\documentclass[letterpaper, 10 pt, conference]{ieeeconf}  

\IEEEoverridecommandlockouts                              

\usepackage{graphics} 
\usepackage{epsfig} 
\usepackage{mathptmx} 
\usepackage{times} 
\usepackage{amsmath} 
\usepackage{amssymb}  

\usepackage{caption}
\usepackage{subcaption}
\usepackage{float}

\newtheorem{definition}{Definition}
\newtheorem{remark}{Remark}

\title{\LARGE \bf
Sliding Mode Control Barrier Function
}

\author{Caio I. G. Chinelato$^{1}$ and Bruno A. Ang\'elico$^{2}$
\thanks{*This work was supported by Funda\c{c}\~ao de Amparo \`a Pesquisa do Estado de S\~ao Paulo (FAPESP) under Grant 2017/22130-4 and Coordenação de Aperfeiçoamento de Pessoal de Nível Superior - Brazil (CAPES) - Finance Code 001}
\thanks{$^{1}$Caio I. G. Chinelato is with Departamento de El\'etrica, Instituto Federal de S\~ao Paulo (IFSP) - Campus S\~ao Paulo, S\~ao Paulo, SP, 01109-010, Brazil, and Departamento de Engenharia de Telecomunica\c{c}\~oes e Controle, Escola Polit\'ecnica da Universidade de S\~ao Paulo (EPUSP), S\~ao Paulo, SP, 05508-010, Brazil
        {\tt\small caio.chinelato@ifsp.edu.br}}%
\thanks{$^{2}$Bruno A. Ang\'elico is with Departamento de Engenharia de Telecomunica\c{c}\~oes e Controle, Escola Polit\'ecnica da Universidade de S\~ao Paulo (EPUSP), S\~ao Paulo, SP, 05508-010, Brazil
        {\tt\small angelico@lac.usp.br}}%
}

\begin{document}

\maketitle
\thispagestyle{empty}
\pagestyle{empty}

\begin{abstract}

This work proposes a sliding mode control barrier function to robustly deal with high relative-degree safety constraints in safety-critical control systems. Stability/tracking objectives, expressed as a nominal control law, and safety constraints, expressed as control barrier functions are unified through quadratic programming. The proposed control framework is numerically validated considering a Furuta pendulum and a magnetic levitation system. For the first system, a linear quadratic regulator is considered as a nominal control law, and a safety constraint is considered to guarantee that the pendulum angular position never exceeds a predetermined value. For the second one, a sliding mode controller is considered as a nominal control law and multiple safety constraints are considered to guarantee that the magnetic levitation system positions never exceed predetermined values. For both systems, we consider high relative-degree safety constraints robust against model uncertainties. The numerical results indicate that the stability/tracking objectives are reached and the safety constraints are respected even with model uncertainties.

\end{abstract}

\section{INTRODUCTION}
\label{sec:introduction}
Safety is a fundamental concept in several engineering problems, such as control systems, robotics and automotive applications. Safety-critical control systems are those that must satisfy stability/tracking objectives and safety constraints, where the safety constraints are prioritized. The safety constraints are specified in terms of a set invariance and verified through control barrier functions (CBFs). The existence of a CBF satisfying specific conditions implies in set invariance. Any trajectory starting inside an invariant set will never reach the complement of the set \cite{Ames_1}.

CBFs can be directly related to control Lyapunov functions (CLFs). CLFs utilize Lyapunov functions together with inequality constraints on the derivatives to establish entire classes of controllers that stabilize a given system \cite{Sontag}, \cite{Freeman}. Several works present applications of CLFs as feedback controllers, such as \cite{CLF_Ames}. Then, the key point is to impose inequality constraints on the derivative of a candidate CBF to establish entire classes of controllers that render a given set forward invariant \cite{Ames_2}.

The unification of CLFs and CBFs can be seen in \cite{Romdlony} and \cite{Ames_3}, using different formulations. In both cases, stability/tracking objectives are expressed as a CLF and safety constraints are expressed as a CBF. The objective of \cite{Romdlony} is to obtain a feedback control law that satisfies simultaneously stability/tracking objectives and safety constraints. The feedback control law is constructed using Sontag's universal control formula \cite{Sontag_1}. Then, the concept of control Lyapunov barrier function is presented. It is important to highlight that if the stability/tracking objectives and the safety constraints are in conflict, then no feedback control law can be proposed. In contrast, \cite{Ames_3} proposes a feedback design problem that mediates stability/tracking objectives and safety constraints, in the sense that safety is always guaranteed. Stability/tracking objectives are satisfied just when the two requirements are not in conflict. In this approach, a quadratic programming (QP) mediates the two inequalities associated with the CLF and CBF. Relaxation is used to make the stability/tracking objectives as a soft constraint and safety as a hard constraint. Then, stability/tracking objectives and safety constraints do not need to be simultaneously satisfiable. Furthermore, in this approach, stability/tracking objectives can be expressed as any linear or nonlinear nominal control law, making the control design more versatile \cite{Ames_1}, \cite{Rauscher_2016}, \cite{Gurriet_2018}.

The approach proposed in \cite{Ames_3} is considered in this work. Several applications using this methodology are proposed in the literature, such as adaptive cruise control (ACC) \cite{Mehra_2015}, lane keeping \cite{Xu_2017}, bipedal walking robot \cite{Nguyen}, robotic manipulator \cite{Rauscher_2016}, robotic grasping \cite{Cortez_2019} two-wheeled human transporter (\textit{Segway}) \cite{Gurriet_2018}, quadrotors \cite{Wu_2016} and multi-robot systems \cite{Wang_2017}.

The control framework described in \cite{Ames_3} is only applicable for safety constraints with relative-degree one, i.e., the first time-derivative of the CBF has to depend on the control input. However, in several systems, such as robotics, it is considered position-based constraints with high relative-degree (greater that one). The works \cite{Nguyen} and \cite{Wu_2016} propose a solution applied only for safety constraints with relative-degree two. In \cite{Hsu}, a backstepping-based method is applied to arbitrary high relative-degree safety constraints. However, this method is challenging and has not been attempted \cite{Ames_1}. In \cite{Nguyen_ECBF}, the concept of exponential control barrier function (ECBF) was first introduced as a way to systematically enforce high relative-degree safety constraints. This work shows that the ECBF can be designed using conventional control design techniques and proposes a pole placement controller for ECBF design.

Furthermore, the control framework described in \cite{Ames_3} does not consider robustness issues, such as disturbances and model uncertainties. Robustness is an essential topic in safety-critical control systems, because if significant disturbances and model uncertainties are not considered, the safety constraints may not be respected. Robust CBFs are studied in \cite{Rob_1}, \cite{Rob_2}, \cite{Rob_3} and \cite{Rob_4}. In \cite{Rob_1}, the robustness of the CBF under model perturbation is investigated. This work verifies set invariance considering perturbations. If the model perturbation has a $H_{\infty }$ norm less than a determined value, input-to-state stability (ISS) property of the set is verified, the set is asymptotic stable and the system is robust. However, this work only describes robustness analysis and do not develop a robust controller design. The design of a robust controller is described in \cite{Rob_2}, considering disturbances, and in \cite{Rob_3} and \cite{Rob_4}, considering model uncertainties.


The main contribution of this paper consists of proposing a sliding mode control barrier function (SMCBF) for systems with high relative-degree safety constraints and model uncertainties. The stability/tracking objectives are expressed as a nominal control law. The proposed control framework is numerically validated considering a Furuta pendulum and a magnetic levitation (MAGLEV) system. For the first system, a linear quadratic regulator (LQR) is considered as a nominal control law and a safety constraint is considered to guarantee that the pendulum angular position never exceeds a predetermined value. For the second one, a sliding mode control (SMC) is considered as a nominal control law and multiple safety constraints are considered to guarantee that the MAGLEV positions never exceed predetermined values. For both systems, we consider high relative-degree safety constraints robust against model uncertainties.

The rest of this paper is organized as follows: In section \ref{sec:CBF}, the concept of CBF and the control framework that unifies the nominal control law and the CBF through QP are presented. The concept of ECBF for safety constraints with high relative-degree and the design of ECBF with pole placement control are described in section \ref{sec:ECBF}. In section \ref{sec:RCBF}, the proposed control framework is described, i.e., the design of SMCBF. The modeling and the numerical results of the Furuta pendulum and the MAGLEV with the proposed control framework are presented in sections \ref{sec:FUR} and \ref{sec:MAG}, respectively. The conclusions are presented in section \ref{sec:CONC}.

\section{CONTROL BARRIER FUNCTION}
\label{sec:CBF}

Given a dynamical system

\begin{equation} \label{eq:SYS}
\dot{x}=f(x)+g(x)u,
\end{equation}
with states $x\in \mathcal{D}\subset \mathbb{R}^{n}$, inputs $u\in \mathcal{U}\subset \mathbb{R}^{m}$ and, $f(x)$ and $g(x)$ locally Lipschitz, there are two types of barrier functions: reciprocal barrier function $B(x)$ and zeroing barrier function $h(x)$ \cite{Ames_2}. Considering a set $\mathcal{C}$ related to the system safety, we have that $B(x)\rightarrow \infty$ as $x\rightarrow \partial \mathcal{C}$ and $h(x)\rightarrow 0$ as $x\rightarrow \partial \mathcal{C}$ \cite{Ames_2}. In this work, we consider $h(x)$. If $B(x)$ or $h(x)$ satisfies Lyapunov-like conditions, then forward invariance of $\mathcal{C}$ is guaranteed \cite{Ames_1}. The natural extension of a barrier function to a system with control inputs is a CBF \cite{Wieland}. As shown for CLFs, in CBFs inequality constraints are imposed on the derivative to obtain entire classes of controllers that render a given set forward invariant.

A feedback controller must be designed for (\ref{eq:SYS}) in order to keep the states $x$ in the safe set $\mathcal{C}$, defined as \cite{Ames_1}:
\begin{equation} \label{eq:SAFE}
\begin{array}{ll}
\mathcal{C}=\left \{ x\in \mathcal{D}\subset \mathbb{R}^{n}:h\left ( x \right )\geq 0 \right \}, \\
\partial \mathcal{C}=\left \{ x\in \mathcal{D}\subset \mathbb{R}^{n}:h\left ( x \right ) = 0 \right \},\\
{\rm Int}(\mathcal{C})=\left \{ x\in \mathcal{D}\subset \mathbb{R}^{n}:h\left ( x \right )>  0 \right \},
\end{array}
\end{equation}
where $h(x):\mathcal{D}\subset \mathbb{R}^{n}\rightarrow \mathbb{R}$ is a continuously differentiable function. $h(x)$ is called a CBF defined on the set $\mathcal{D}$ with $\mathcal{C}\subseteq \mathcal{D}\subset \mathbb{R}^{n}$, if there exists an extended class $\kappa $ functions $\alpha_{cbf} $ such that \cite{Ames_2}
\begin{equation} \label{eq:CBF1}
\sup_{u\in \mathcal{U}}\left [L_{f}h(x)+L_{g}h(x)u +\alpha_{cbf}(h(x)) \right ]\geq 0,
\end{equation}
where $L_{f}h=\nabla h(x) \cdot f(x)$ and $L_{g}h=\nabla h(x) \cdot g(x)$.

\begin{remark}   
A continuous function $\alpha_{cbf}:[0,a)\rightarrow[0,\infty )$ for some $a> 0$ is said to belong to class $\kappa $ if it is strictly increasing and $\alpha_{cbf}(0)=0$ \cite{Ames_2}.
\end{remark}

Considering the safe set $\mathcal{C}$ defined by (\ref{eq:SAFE}), the CBF $h(x)$ for the system (\ref{eq:SYS}) and the set
\begin{equation} \label{eq:CBF2}
K_{cbf}(x)=\left \{ u\in \mathcal{U}:L_{f}h(x)+L_{g}h(x)u +\alpha_{cbf}(h(x))\geq 0 \right \},
\end{equation}
the work \cite{Ames_2} enunciates that any locally Lipschitz continuous controller $u:\mathcal{D}\rightarrow \mathcal{U}$ such that $u(x)\in K_{cbf}(x)$ will render the set $\mathcal{C}$ forward invariant.

The final control framework unifies stability/tracking objectives, expressed as a nominal control law, and safety constraints, expressed as a CBF, through QP. The controller is formulated as an optimization problem, minimizing the error \cite{Rauscher_2016}
\begin{equation} \label{eq:CBF3}
e_{u}=u_{no}-u,
\end{equation}
where $u_{no}$ is the nominal controller for the system (\ref{eq:SYS}).

The squared norm of the error
\begin{equation} \label{eq:CBF4}
\left \| e_{u} \right \|^{2}=u^{T}u-2u_{no}^{T}u+u_{no}^{T}u_{no}
\end{equation}
is considered as the objective function. The last term of (\ref{eq:CBF4}) is neglected, since it is constant in a minimization process with respect to $u$. Thus, we can consider the following QP-based controller \cite{Ames_1}, \cite{Rauscher_2016}:
\begin{equation} \label{eq:CBF5}
\begin{array}{ll}
u^{*}=\underset{u\in {\mathbb{R}}^{m}}{\arg\min}\;u^{T}u-2u_{no}^{T}u\\
s.t.\;\;L_{g}h(x)u+L_{f}h(x)+\alpha_{cbf} (h(x))\geq 0.
\end{array}
\end{equation}

It is important to highlight that the constraint in QP enforces the condition (\ref{eq:CBF1}) for CBF.

The QP-based controller (\ref{eq:CBF5}) is only applicable for safety constraints with relative-degree one, i.e., the first time-derivative of the CBF has to depend on the control input. In this work, we consider high relative-degree safety constraints. In this case, as $L_{g}h(x)=0$, the QP cannot be solved. One way to systematically enforce high relative-degree safety constraints is the application of ECBF. Furthermore, the controller does not consider robustness issues, such as model uncertainties; thus, the next sections describe the concept of ECBF, the design of ECBF with pole placement control and the design of SMCBF to deal with model uncertainties, which is the main contribution of this paper.

\section{EXPONENTIAL CONTROL BARRIER FUNCTION WITH POLE PLACEMENT CONTROL}
\label{sec:ECBF}

The concept of ECBF was first introduced in \cite{Nguyen_ECBF}, where in the final control framework, stability/tracking objectives are expressed as a CLF and the ECBF is derived based on the reciprocal CBF $B(x)$. However, in this work, the formulation of \cite{Nguyen_ECBF} is adapted in order to express stability/tracking objectives as a nominal control law, such as in (\ref{eq:CBF5}), and the ECBF is derived based on the zeroing CBF $h(x)$, such as in \cite{Ames_1}.

The term ECBF is used since the resulting CBF constraint is an exponential function of the initial condition. Furthermore, the design of ECBFs is based on the linear control theory, so conventional linear control design techniques such as pole placement can be applied \cite{Nguyen_ECBF}.

\begin{definition}   
Given a set $\mathcal{C}\subset \mathcal{D}\subset \mathbb{R}^{n}$ defined as the superlevel set of a $r$-times continuously differentiable function $h(x):\mathcal{D}\rightarrow \mathbb{R}$, then $h(x)$ is an ECBF if there exists a row vector $K_{b}\in \mathbb{R}^{r}$ such that for the control system (\ref{eq:SYS}) \cite{Ames_1},
\begin{equation} \label{eq:ECBF1}
\sup_{u\in \mathcal{U}}\left [L_{f}^{r}h(x)+L_{g}L_{f}^{r-1}h(x)u+K_{b}\eta _{b}(x) \right ]\geq 0
\end{equation}
$\forall x\in {\rm Int}(\mathcal{C})$ results in $h(x(t))\geq C_{b}e^{A_{b}t}\eta _{b}(x_{0})\geq 0$, whenever $h(x_{0})\geq 0$, where the matrix $A_{b}$ is dependent on the choice of $K_{b}$, and
\begin{equation} \label{eq:ECBF2}
\eta _{b}(x):= \left [ \begin{array}{ll}h(x)\\\dot{h}(x)\\\ddot{h}(x) \\ \vdots \\h^{(r-1)}(x) \\\end{array} \right ]=\left [ \begin{array}{ll}h(x)\\L_{f}h(x)\\L_{f}^{2}h(x) \\ \vdots \\L_{f}^{(r-1)}h(x) \\\end{array} \right ],
\end{equation}
\begin{equation} \label{eq:ECBF3}
C_{b}=\left [ \begin{array}{ll}1\;\;0\;\;\cdots\;\;0 \\\end{array} \right ].
\end{equation}
\end{definition}

In \cite{CLF_Ames}, it is described a systematic procedure using input-output linearization to design CLFs for regulating outputs with arbitrary relative-degree. This procedure could be applied to design CBFs for constraints with arbitrary relative-degree $r$. However, this procedure is not directly feasible to $\dot{h}(x,u)=L_{f}h(x)+L_{g}h(x)u$ since $L_{g}h(x)$ is a vector and obviously not invertible \cite{Nguyen_ECBF}. The work \cite{Nguyen_ECBF} introduces the notion of
virtual input-ouput linearization (VIOL) wherein an invertible decoupling matrix is not required. Considering a virtual control input $\mu _{b}$ defined as
\begin{equation} \label{eq:ECBF4}
h^{(r)}(x,u)=L_{f}^{r}h(x)+L_{g}L_{f}^{r-1}h(x)u:=\mu _{b},
\end{equation}
such that the input-output linearized system becomes
\begin{equation} \label{eq:ECBF5}
\begin{array}{ll}\dot{\eta }_{b}(x)=F_{b}\eta_{b}(x)+G_{b}\mu_{b},\\ h(x)=C_{b}\eta_{b}(x),  \\\end{array}
\end{equation}
where $\eta_{b}(x)$ is defined in (\ref{eq:ECBF2}), $F_{b}\in \mathbb{R}^{r\times r}$ and $G_{b}\in \mathbb{R}^{r\times 1}$ are defined as
\begin{equation} \label{eq:ECBF6}
F_{b} = \left [ \begin{array}{ll}0\;\;1\;\;0\;\;\cdots \;\;0  \\ 0\;\;0\;\;1\;\;\cdots \;\;0  \\ \vdots \;\;\;\vdots \;\;\;\vdots \;\;\;\ddots \;\;\vdots \\ 0\;\;0\;\;0\;\;\cdots \;\;1    \\  0\;\;0\;\;0\;\;\cdots \;\;0 \\\end{array} \right ],G_{b}=\left [ \begin{array}{ll}0\\0\\\vdots \\0\\1 \\\end{array} \right ],
\end{equation}
and $C_{b}$ is as defined in (\ref{eq:ECBF3}).

If we want to drive $h(x)$ to zero, \cite{Nguyen_ECBF} proposes to design the EBCF with a pole placement controller, $\mu _{b}=-K_{b}\eta_{b}$, with all negative real poles $p_{b}=-\left [ \begin{array}{ll}p_{1}\;\;p_{2}\;\;\cdots\;\;p_{r} \\\end{array} \right ]$, where $p_{i}>0,\;i=1,\cdots ,r$; thus, $h(x(t))= C_{b}e^{A_{b}t}\eta _{b}(x_{0})$, where the closed-loop matrix $A_{b}=F_{b}-G_{b}K_{b}$ with all negative real eigenvalues.

Similarly to (\ref{eq:CBF5}), a nominal control law $u_{no}$ and an ECBF $h(x)$ can be unified using QP considering the following controller \cite{Ames_1}, \cite{Nguyen_ECBF}:
\begin{equation} \label{eq:ECBF7}
\begin{array}{ll}
u^{*}(x)=\underset{u=(u,\mu_{b})\in {\mathbb{R}}^{m+1}}{\arg\min}\;u^{T}u-2u_{no}^{T}u\\
s.t.\;\;L_{g}L_{f}^{r-1}h(x)u+L_{f}^{r}h(x)=\mu_{b}, \\
\;\;\;\;\;\;\;\mu_{b}\geq  -K_{b}\eta _{b}(x). \\
\end{array}
\end{equation}


\section{SLIDING MODE CONTROL BARRIER FUNCTION}
\label{sec:RCBF}

Considering the virtual input-output linearized system (\ref{eq:ECBF5}), the basic idea is to design the CBF with a sliding mode controller $\mu_{b}$ in order to deal with model uncertainties, instead of applying a pole placement controller as described in the ECBF (\ref{eq:ECBF7}). It is considered that $f(x)$ and $g(x)$ in (\ref{eq:SYS}) represent the real dynamics and are unknown. However, the controller's design is based on the nominal dynamics $\bar{f}(x)$ and $\bar{g}(x)$; thus, the effect of model uncertainties in VIOL (\ref{eq:ECBF4}) can be described such as

\begin{equation} \label{eq:RCBF1}
h^{(r)}(x,\Delta,\mu_{b})=\mu_{b}+\Delta,
\end{equation}
where $\Delta$ is related to the difference between the real dynamics and the nominal dynamics in the Lie derivatives of (\ref{eq:ECBF4}), i.e., the model uncertainty. We assume that $\Delta$ is bounded, i.e.,
\begin{equation} \label{eq:RCBF2}
\left \| \Delta \right \|\leq \Delta _{\max}.
\end{equation}

Considering a time-varying surface $S(x,t)$ in state-space $\mathbb{R}^{n}$ defined by
\begin{equation} \label{eq:RCBF3}
S(x,t)=\left ( \frac{d}{dt}+\lambda  \right )^{r-1}\tilde{h},
\end{equation}
where $\tilde{h}=h-h_{d}$ and $\lambda$ is a strictly positive constant \cite{Slotine}. We consider the CBF desired value $h_{d}\geq0$ since we want to guarantee that $h\geq 0$. The problem of tracking $h=h_{d}$ is equivalent to that of remaining on the surface $S$ for all $t>0$. Furthermore, $S=0$ represents a linear differential equation whose unique solution is $\tilde{h}=0$; thus, the problem of tracking $h_{d}$ can be reduced to that of keeping the scalar quantity $S$ to zero \cite{Khalil}.

The problem of keeping $S$ at zero can be achieved by choosing a control law $\mu_{b}$ such that $S$ satisfies

\begin{equation} \label{eq:RCBF4}
\frac{1}{2}\frac{d}{dt}S^{2}\leq -\eta \left | S \right |,
\end{equation}
being $\eta$ a strictly positive constant \cite{Slotine}. The condition (\ref{eq:RCBF4}), called sliding condition, demonstrates that, once on the surface, the system trajectories remains it, i.e., the surface is an invariant set. Therefore, model uncertainties and disturbances can be tolerated. $S$ is denominated sliding surface, and the system's behaviour once on the surface is called sliding mode \cite{Slotine}.

In the SMC design, a feedback control law, called equivalent control, is determined to maintain the system in sliding mode, i.e., $\dot{S}=0$. However, in order to deal with model uncertainties and disturbances, the control law has to be discontinuous across $S$ \cite{Slotine}.

In this work, we consider safety constraints with relative-degree $r=2$; thus, the sliding surface $S$, defined by (\ref{eq:RCBF3}), is given by
\begin{equation} \label{eq:RCBF5}
S=\dot{\tilde{h}}+\lambda\tilde{h},
\end{equation}
and using (\ref{eq:RCBF1}),
\begin{equation} \label{eq:RCBF6}
\dot{S}=\ddot{h}-\ddot{h}_{d}+\lambda\dot{\tilde{h}}=\mu_{b}+\Delta-\ddot{h}_{d}+\lambda\dot{\tilde{h}}.
\end{equation}

The equivalent control $\bar{\mu}_{b}$ designed using the nominal dynamics and that would achieve $\dot{S}=0$ is given by
\begin{equation} \label{eq:RCBF7}
\bar{\mu}_{b} = \ddot{h}_{d}-\lambda\dot{\tilde{h}}.
\end{equation}

Considering $h(x_0) \geq 0$, i.e., that $x_0$ is within the safe set, the control law given by
\begin{equation} \label{eq:RCBF8}
\mu_{b}=\bar{\mu}_{b}-K_{smc} \, {\rm sgn}(S),
\end{equation}
where ${\rm sgn}$ is the sign function, will drive $h$ to $h_d$ despite the bounded uncertainty $\Delta$ in \eqref{eq:RCBF1}-\eqref{eq:RCBF2}. Using (\ref{eq:RCBF4}) and (\ref{eq:RCBF8}), we obtain
\begin{equation} \label{eq:RCBF9}
\frac{1}{2}\frac{d}{dt}S^{2}=\dot{S}S=\Delta S-K_{smc}\left | S \right |\leq -\eta \left | S \right |,
\end{equation}
and $K_{smc}$ that satisfies the sliding mode condition (\ref{eq:RCBF4}) is then given by
\begin{equation} \label{eq:RCBF10}
K_{smc}\geq \Delta +\eta.
\end{equation}

The discontinuous term in (\ref{eq:RCBF8}) generate a control switching that is necessarily imperfect, because switching is not instantaneous, and the value of $S$ is not known with infinite precision and is never exactly zero. This can lead to chattering, i.e., undesirable high-frequency actuation and vibration. To avoid chattering, it is applied a boundary layer in the neighboring of the sliding surface and a saturation function replaces the sign function \cite{Slotine}, \cite{Khalil}, such as
\begin{equation} \label{eq:RCBF11}
\mu_{b}=\bar{\mu}_{b}-K_{smc} \, {\rm sat}(S/\Phi ),
\end{equation}
where
\begin{equation} \label{eq:RCBF12}
{\rm sat}(S/\Phi )=\left\{\begin{matrix}
S/\Phi,\;{\rm if}\;\left | S \right |\leq\Phi \\
{\rm sgn}(S/\Phi ),\;{\rm if}\;\left | S \right |> \Phi,
\end{matrix}\right.
\end{equation}
and $\Phi$ is the boundary layer thickness. Without the boundary layer, a reasonable choice would be $h_d =0$. However, with the boundary layer, $h_d$ has to be somewhat greater than zero to guarantee that $h \geq 0$.

Then, for the SMCBF, the QP-based controller can be described by
\begin{equation} \label{eq:RCBF13}
\begin{array}{ll}
u^{*}(x)=\underset{u=(u,\mu_{b})\in {\mathbb{R}}^{m+1}}{\arg\min}\;u^{T}u-2u_{no}^{T}u\\
s.t.\;\;L_{g}L_{f}^{r-1}h(x)u+L_{f}^{r}h(x)=\mu_{b} \\
\;\;\;\;\;\;\;\mu_{b}\geq  \bar{\mu}_{b}-K_{smc}\, {\rm sat}(S/\Phi ). \\
\end{array}
\end{equation}

\section{FURUTA PENDULUM}
\label{sec:FUR}

This section presents the modeling and the numerical results of the Furuta pendulum with the proposed control framework.

\subsection{System Modeling}

\begin{figure}[h]
\begin{center}
\includegraphics[width=4.5cm]{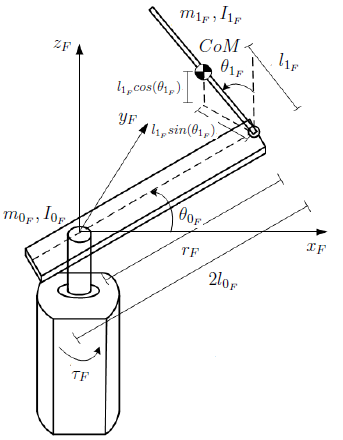}
\caption{Schematic diagram of the Furuta pendulum \cite{FUR_1}.}
\label{fig:FUR_SYS}
\end{center}
\end{figure}

The schematic diagram of the Furuta pendulum is shown in Fig. \ref{fig:FUR_SYS} \cite{FUR_1}, \cite{FUR_2}. The pendulum parameters are the rotating arm mass $m_{0_{F}}$, the pendulum mass $m_{1_{F}}$, the rotating arm length $2l_{0_{F}}$, the pendulum length $2l_{1_{F}}$, the distance from the fixed axis to the pendulum basis $r_F$ and the center of mass (CoM) of the pendulum arm $d_F$ w.r.t fixed axis \cite{FUR_1}.

The system dynamics are defined using the Euler-Lagrange formulation:
\begin{equation} \label{eq:FUR1}
\frac{d}{dt}\left ( \frac{\partial L_{F}}{\partial \dot{\theta}_{i_{F}}} \right )-\frac{\partial L_{F}}{\partial \theta_{i_{F}}}=Q _{i_{F}},\;\;i=0,1
\end{equation}
where $L_{F}=K_{0_{F}}+K_{1_{F}}-P_{F}$ is the system Lagrangian, $K_{0_{F}}$ is the kinetic energy of the rotating arm, $K_{1_{F}}$ is the kinetic energy of the pendulum, $P_{F}$ is the total potential energy, $\theta _{0_{F}}$ and $\theta _{1_{F}}$ are generalized coordinates and $Q _{0_{F}}$ and $Q _{1_{F}}$ are the generalized forces (torques).

The potential energy $P_{F}$ can be described using the displacement of the CoM of the pendulum \cite{FUR_1}, such as
\begin{equation} \label{eq:FUR2}
P_{F}=m_{1_{F}}gl_{1_{F}}cos(\theta _{1_{F}}).
\end{equation}

The kinetic energy of the rotating arm $K_{0_{F}}$ is composed only by the rotation \cite{FUR_1}. Hence,
\begin{equation} \label{eq:FUR3}
K_{0_{F}}=\frac{I_{0_{F}}\dot{\theta }_{0_{F}}^{2}}{2},
\end{equation}
where the inertia $I_{0_{F}}$ is given by
\begin{equation} \label{eq:FUR31}
I_{0_{F}}=\frac{m_{0_{F}}(2l_{0_{F}})^{2}}{12}+m_{0_{F}}d_{F}^{2}.
\end{equation}

The kinetic energy of the pendulum $K_{1_{F}}$ can be described using its rotation, the velocity of the CoM in the $x_{F}$-direction and in $y_{F}$-direction \cite{FUR_1}, such as
\begin{equation} \label{eq:FUR4}
\begin{array}{ll}
K_{1_{F}}=\frac{I_{1_{F}}\dot{\theta }_{1_{F}}^{2}}{2}+\frac{m_{1_{F}}}{2} (l_{1_{F}}^{2}\dot{\theta }_{1_{F}}^{2} + r_{F}^{2}\dot{\theta }_{0_{F}}^{2}\\+l_{1_{F}}^{2}\dot{\theta }_{0_{F}}^{2}sin^{2}(\theta _{1_{F}})+ 2r_{F}l_{1_{F}}\dot{\theta }_{0_{F}}\dot{\theta }_{1_{F}}cos(\theta _{1_{F}}) ),
\end{array}
\end{equation}
where the inertia $I_{1_{F}}$ is given by
\begin{equation} \label{eq:FUR41}
\begin{array}{ll}
I_{1_{F}}=\frac{m_{1_{F}}(2l_{1_{F}})^{2}}{12}.
\end{array}
\end{equation}

The torque of the DC motor $\tau _{F}$ is proportional to the duty-cycle of the Pulse Width Modulation (PWM) signal $V_{m_{F}}\in \left [ -1,1 \right ]$ \cite{FUR_1} and can be expressed as \footnote{Actually, $V_{m_{F}}$ includes the PWM duty-cycle and the direction of the motor rotation, being (+) counterclockwise and (-) clockwise.}
\begin{equation} \label{eq:FUR5}
\tau _{F}=\frac{K_{t_{F}}}{R_{m_{F}}}(2V_{m_{F}}-K_{e_{F}}\dot{\theta }_{0_{F}}),
\end{equation}
where $K_{t_{F}}$ is the motor torque constant, $R_{m_{F}}$ is the armature resistance and $K_{e_{F}}$ is the back EMF constant.

Lastly, the generalized forces $Q _{0_{F}}$ and $Q _{1_{F}}$ of (\ref{eq:FUR1}) are defined as external forces and reaction forces with relation to each generalized variable $\theta _{0_{F}}$ and $\theta _{1_{F}}$:
\begin{equation} \label{eq:FUR6}
Q _{0_{F}}=\tau _{F}-b _{0_{F}}\dot{\theta }_{0_{F}},
\end{equation}
\begin{equation} \label{eq:FUR7}
Q _{1_{F}}=-b _{1_{F}}\dot{\theta }_{1_{F}}.
\end{equation}

For the rotating arm, there is a torque applied by the DC motor and there is a reaction torque due to viscous damping of motor shaft and gearbox ($b _{0_{F}}$). For the pendulum, there is a reaction torque due to viscous damping of pendulum bearing and encoder coupling ($b _{1_{F}}$) \cite{FUR_1}.

In order to design the linear control to stabilize the system, the nonlinear model is linearized around the equilibrium point, resulting in:
\begin{equation} \label{eq:FUR8}
\dot{x}_{F}=Ax_{F}+Bu_{F},
\end{equation}
where $x_{F} = \left [ \theta _{0_{F}} \;\;\theta _{1_{F}}\;\;\dot{\theta }_{0_{F}}\;\;\dot{\theta }_{1_{F}}  \right]^{T}$, $u_{F}=V_{m_{F}}$, $A$ is the state matrix, $B$ is the input matrix and the equilibrium point is $x^{*}_{F} = \left [ 0 \;\;0\;\;0\;\;0  \right]^{T}$. It is important to highlight that (\ref{eq:FUR8}) can be related to (\ref{eq:SYS}), where $f(x)=Ax_{F}$ and $g(x)=B$.

\subsection{Nominal Control - LQR}

LQR is applied as a nominal control law $u_{no_{F}}$ for stabilizing the pendulum. LQR is an optimal regulator that, given the system equation (\ref{eq:FUR8}), determines the matrix $K_{lqr}$ of the optimal control vector
\begin{equation} \label{eq:LQR1}
u_{no_{F}}=-K_{lqr}x_{F}
\end{equation}
so as to minimize the performance index
\begin{equation} \label{eq:LQR2}
J = \int_{0}^{\infty }\left ( x_{F}^{T}Qx_{F}+u_{no_{F}}^{T}Ru_{no_{F}} \right ){\rm d}t,
\end{equation}
where $Q$ is a positive-semidefinite matrix and $R$ is a positive-definite matrix. These matrices are selected to weight the relative importance of the state vector $x_{F}$ and the input $u_{no_{F}}$ on the performance index minimization \cite{Ogata}.

If there exists a positive-definite matrix $P$ satisfying the Riccati equation
\begin{equation} \label{eq:LQR3}
A^{T}P+PA-PBR^{-1}B^{T}P+Q=0,
\end{equation}
then the closed-loop system is stable. Thus, the optimal matrix $K_{lqr}$ can be obtained by
\begin{equation} \label{eq:LQR4}
K_{lqr}=R^{-1}B^{T}P.
\end{equation}

\subsection{Numerical Results}

The behavior of the Furuta pendulum with the proposed control framework is verified through numerical simulations with MATLAB/Simulink. The numerical values of the parameters are $m_{0_{F}}=0.393\;\rm Kg$, $m_{1_{F}}=0.068\;\rm Kg$, $2l_{0_{F}}=0.365\;\rm m$, $2l_{1_{F}}=0.207\;\rm m$, $r_{F}=0.210\;\rm m$, $d_{F}=0.022\;\rm m$, $g=9.81\;\rm m/s^{2}$, $K_{t_{F}}=0.02\;\rm {Nm}/{A}$, $K_{e_{F}}=0.08\;\rm {Vs}/{rad}$ and $R_{m_{F}}=2.4\;\rm \Omega$.

Several works are proposed to satisfy a stability objective in the Furuta pendulum, i.e, to stabilize the system at the equilibrium point, but safety constraints are not considered. So, it is applied the control framework described in this work to simultaneously satisfy stability objectives and safety constraints. LQR is applied as a nominal control law $u_{no_{F}}$, defined in (\ref{eq:LQR1}), for stabilizing the pendulum. The safety constraint is considered to guarantee that the pendulum angular position $\left | \theta _{1_{F}} \right |$ never exceeds a predetermined value $\theta _{1_{F\max}}=0.087\;\rm rad$ $(5^\circ)$. The rotating arm mass $m_{0_{F}}$ and the pendulum mass $m_{1_{F}}$ are increased 60\% to verify the control framework robustness.

The linearized nominal model in (\ref{eq:FUR8}) has the following numerical matrices
\begin{equation} \label{eq:FURES1}
A = \left [ \begin{array}{cccc}0 & 0 & 1 & 0\\0 & 0 & 0 & 1\\0 & -19.8123 & -0.1446 & 0.0003\\0 & 101.2361 & 0.2200 & -0.0015 \end{array} \right ],
\end{equation}

\begin{equation} \label{eq:FURES11}
B = \left [ \begin{array}{cccc}0\\0\\18.8571\\-28.6956 \end{array} \right ].
\end{equation}

For the LQR, we considered
\begin{equation} \label{eq:FURES2}
Q = \left [ \begin{array}{cccc}500 & 0 & 0 & 0\\0 & 500 & 0 & 0\\0 & 0 & 500 & 0\\0 & 0 & 0 & 500 \end{array} \right ],\;R = 1,
\end{equation}
resulting in
\begin{equation} \label{eq:FURES3}
K_{lqr} = \left [ \begin{array}{cccc}-22.3607&-341.0460&-28.6975&-46.0316\\ \end{array} \right ].
\end{equation}

\begin{figure}[h]
     \centering
     \begin{subfigure}[b]{0.23\textwidth}
         \centering
         \includegraphics[trim=10 5 40 15, clip,width=\textwidth]{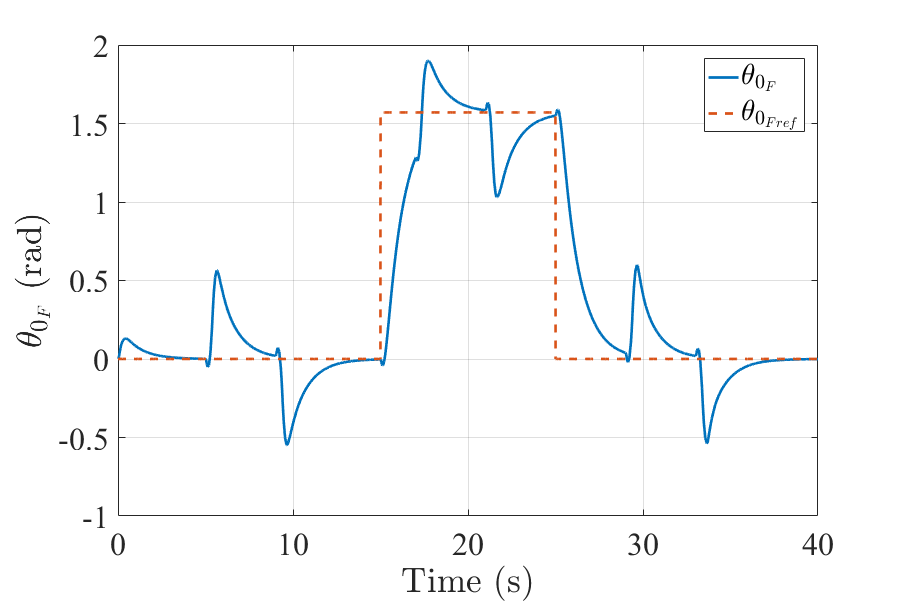}
         \caption{}
     \end{subfigure}
     \begin{subfigure}[b]{0.23\textwidth}
         \centering
         \includegraphics[trim=10 5 40 15, clip,width=\textwidth]{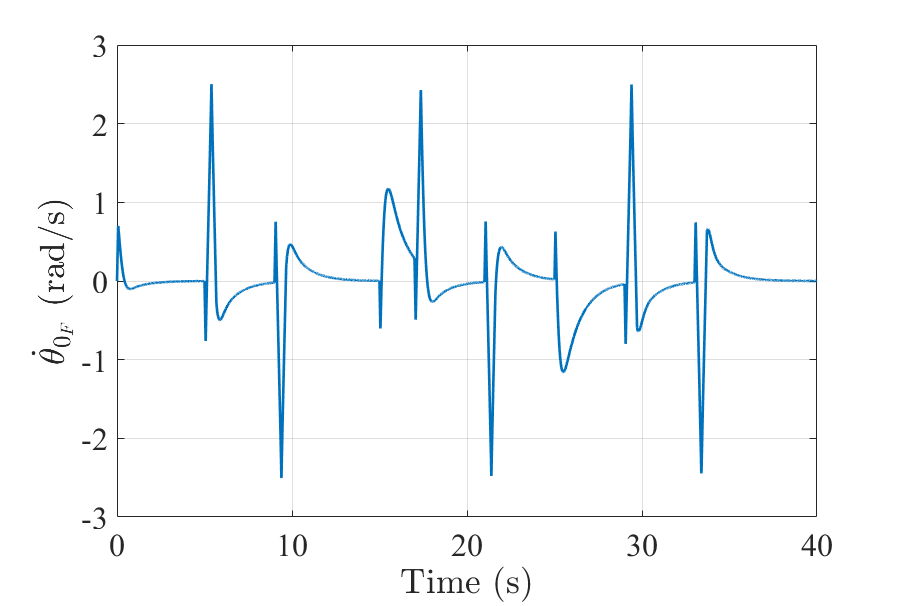}
         \caption{}
     \end{subfigure}
     \begin{subfigure}[b]{0.23\textwidth}
         \centering
         \includegraphics[trim=3 5 47 15, clip,width=\textwidth]{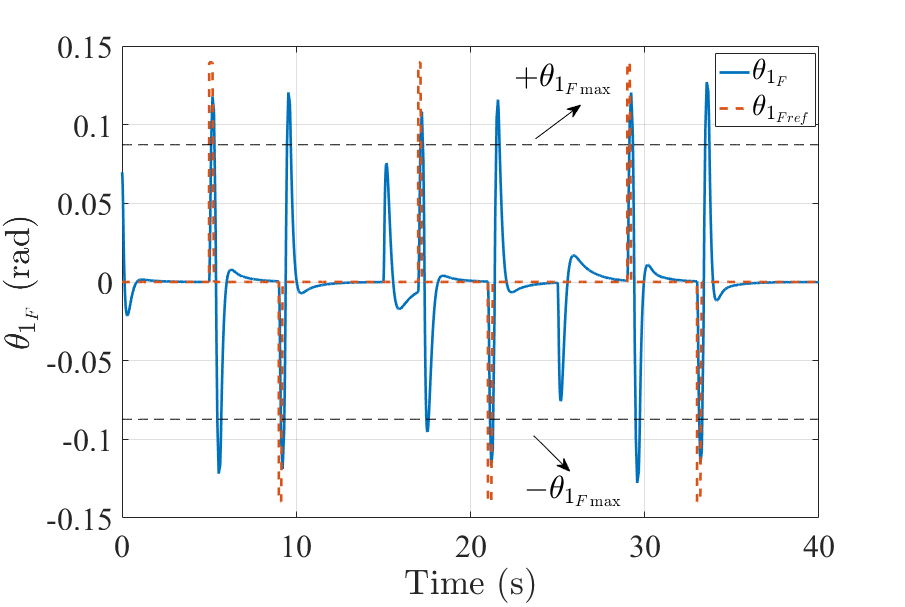}
         \caption{}
    \end{subfigure}
         \begin{subfigure}[b]{0.23\textwidth}
         \centering
         \includegraphics[trim=10 5 40 15, clip,width=\textwidth]{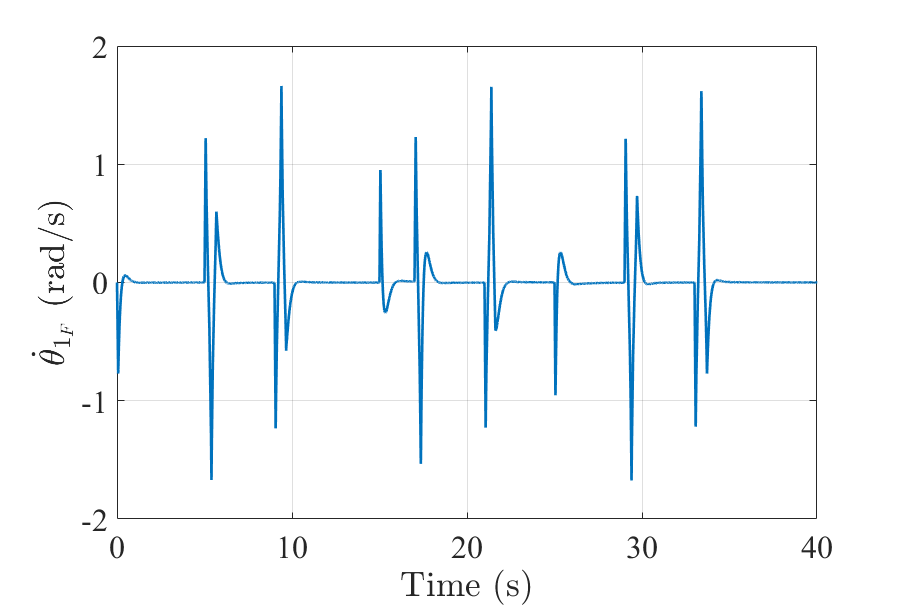}
         \caption{}
    \end{subfigure}
           \begin{subfigure}[b]{0.23\textwidth}
         \centering
         \includegraphics[trim=10 5 40 15, clip,width=\textwidth]{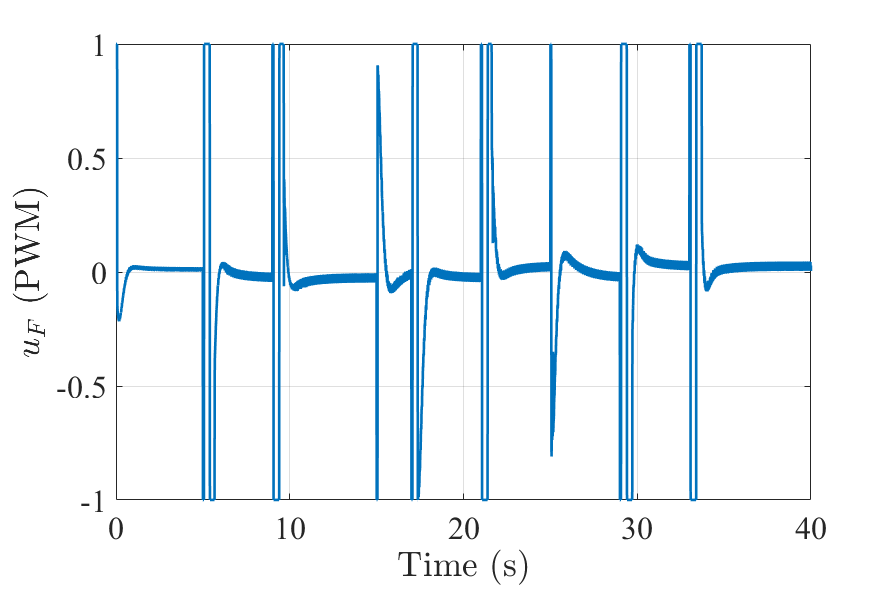}
         \caption{}
    \end{subfigure}
        \caption{Numerical simulation (Furuta pendulum) - LQR without ECBF.}
        \label{fig:FUR_SIM_1}
\end{figure}

It is proposed an experiment whereby the rotating arm angle $\theta _{0_{F}}$ should track a square wave signal reference input $\theta _{0_{Fref}}$ and the pendulum angle $\theta _{1_{F}}$ should track a reference input $\theta _{1_{Fref}}$ composed of short-time pulses. This is considered in order to verify the effect of the CBF, i.e., with the final control framework, $\theta _{1_{F}}$ is expected not to exit the safe set (\ref{eq:SAFE}). Initially, only the LQR is applied and the safety constraint is not considered. When a reference input is considered, the control input (\ref{eq:LQR1}) becomes
\begin{equation} \label{eq:FURES4}
u_{no_{F}}=-K_{lqr}x_{F}+k_{1_{lqr}}\theta _{0_{Fref}}+k_{2_{lqr}}\theta _{1_{Fref}},
\end{equation}
where $k_{1_{lqr}}=-22.3607$ and $k_{2_{lqr}}=-341.0460$. The simulation results are presented in Fig. \ref{fig:FUR_SIM_1}. In all numerical simulations, the pendulum is assumed to start at an initial angular position $\theta _{1_{ini}} = 0.069\;\rm rad$ $(4^\circ)$. The results show that the LQR is able to stabilize the system even with increasing in the masses $m_{0_{F}}$ and $m_{1_{F}}$.

Posteriorly, the safety constraint is considered to guarantee that $\left | \theta _{1_{F}} \right |$ never exceeds $\theta _{1_{F\max}}=0.087\; \rm rad$ $(5^\circ)$ and being always inside the safe set (\ref{eq:SAFE}); thus, the QP-based controller (\ref{eq:ECBF7}) that unifies the nominal LQR $u_{no_{F}}$ and the safety constraint is applied. We consider the following relative-degree two ($r=2$) safety constraint, expressed as the CBF
\begin{equation} \label{eq:FURES5}
h_{F}(x_{F})=\theta _{1_{F\max}}^{2}-\theta _{1_{F}}^{2}.
\end{equation}

\begin{figure}[h]
     \centering
     \begin{subfigure}[b]{0.23\textwidth}
         \centering
         \includegraphics[trim=10 5 40 15, clip,width=\textwidth]{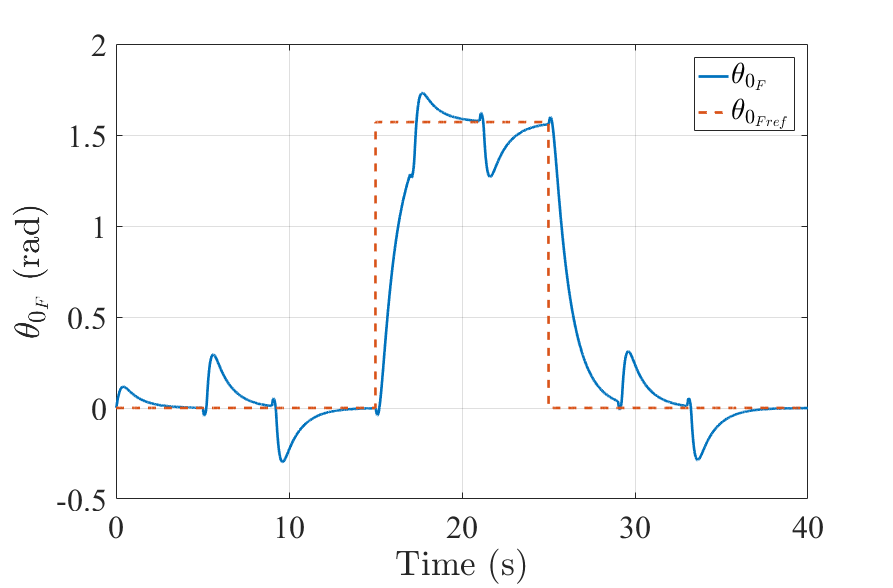}
         \caption{}
     \end{subfigure}
     \begin{subfigure}[b]{0.23\textwidth}
         \centering
         \includegraphics[trim=10 5 40 15, clip,width=\textwidth]{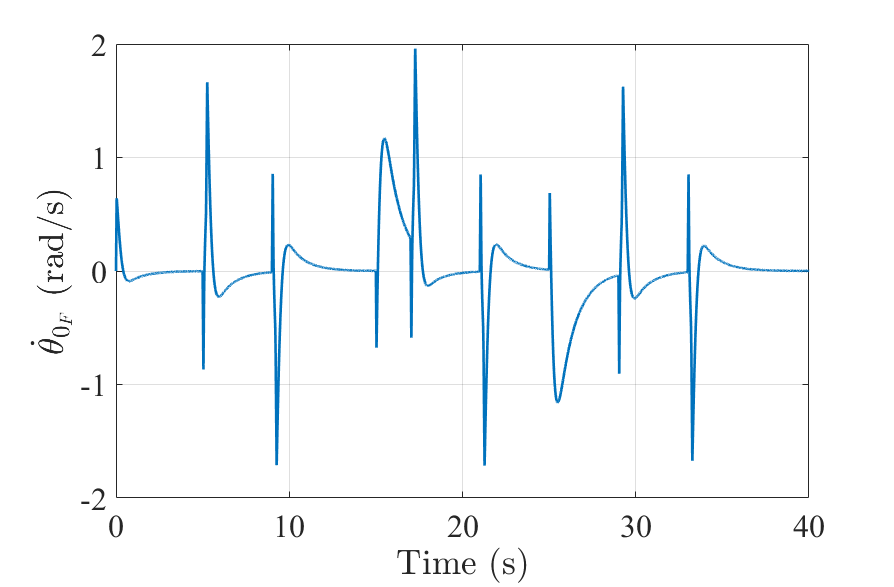}
         \caption{}
     \end{subfigure}
     \begin{subfigure}[b]{0.23\textwidth}
         \centering
         \includegraphics[trim=3 5 47 15, clip,width=\textwidth]{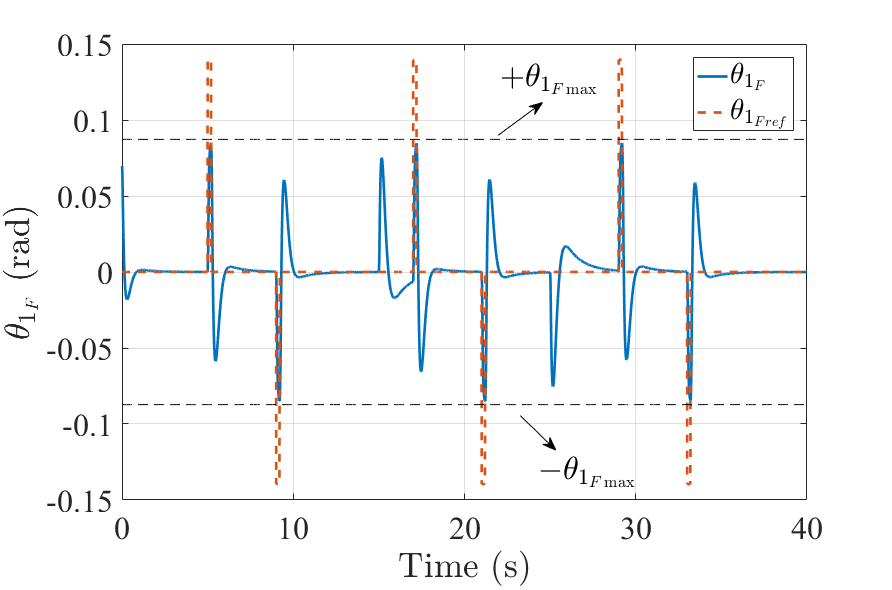}
         \caption{}
    \end{subfigure}
         \begin{subfigure}[b]{0.23\textwidth}
         \centering
         \includegraphics[trim=3 5 47 15, clip,width=\textwidth]{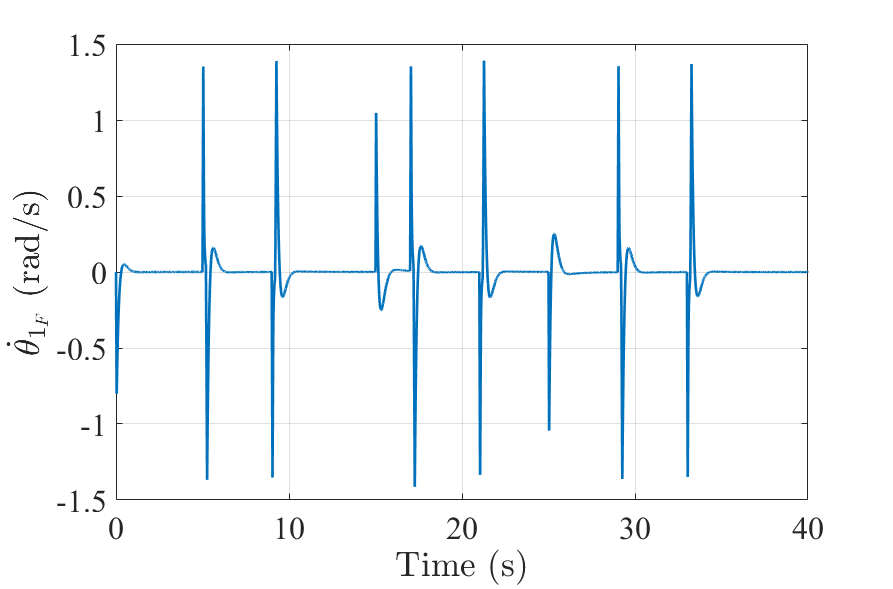}
         \caption{}
    \end{subfigure}
         \begin{subfigure}[b]{0.23\textwidth}
         \centering
         \includegraphics[trim=10 5 40 15, clip,width=\textwidth]{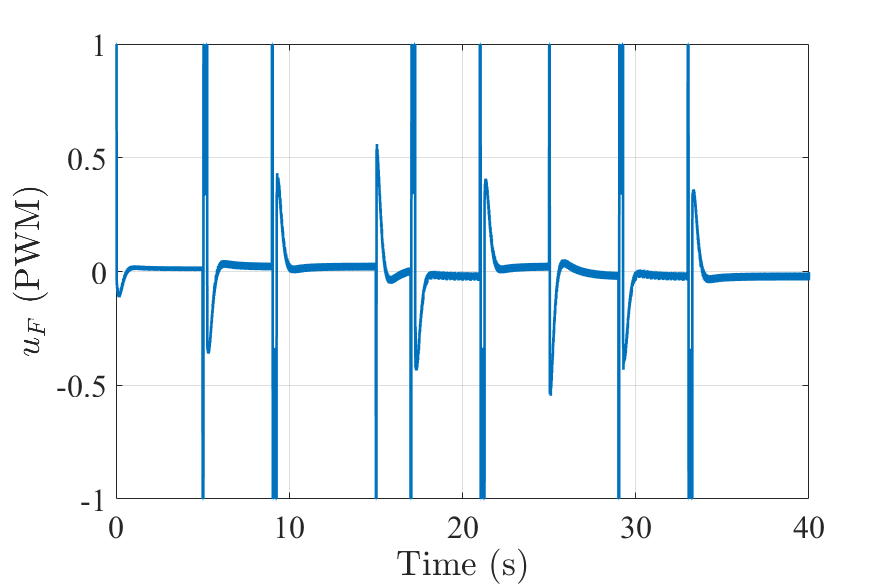}
         \caption{}
    \end{subfigure}
            \begin{subfigure}[b]{0.23\textwidth}
         \centering
         \includegraphics[trim=10 5 40 5, clip,width=\textwidth]{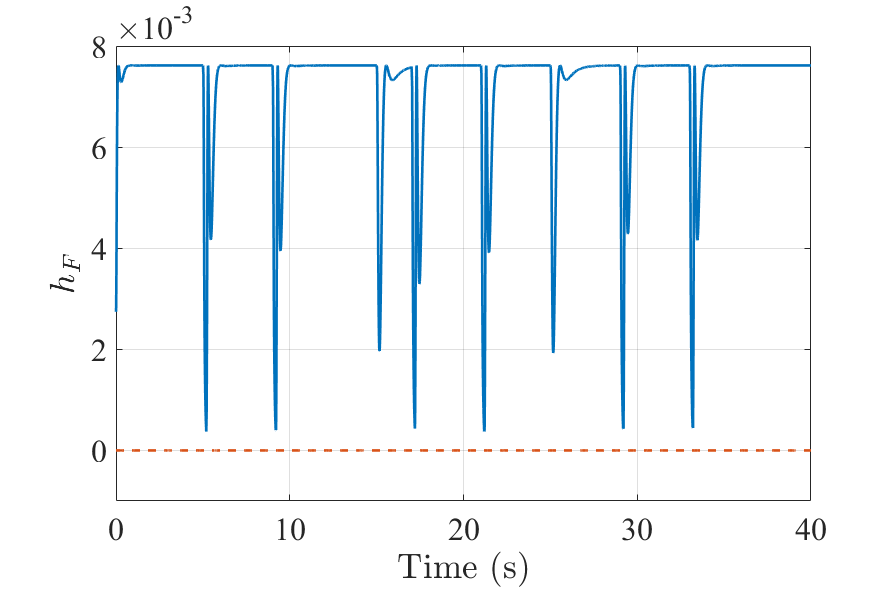}
         \caption{}
    \end{subfigure}
        \caption{Numerical simulation (Furuta pendulum) - LQR with ECBF considering nominal dynamics.}
        \label{fig:FUR_SIM_2}
\end{figure}

\begin{figure}[h]
     \centering
     \begin{subfigure}[b]{0.23\textwidth}
         \centering
         \includegraphics[trim=10 5 40 15, clip,width=\textwidth]{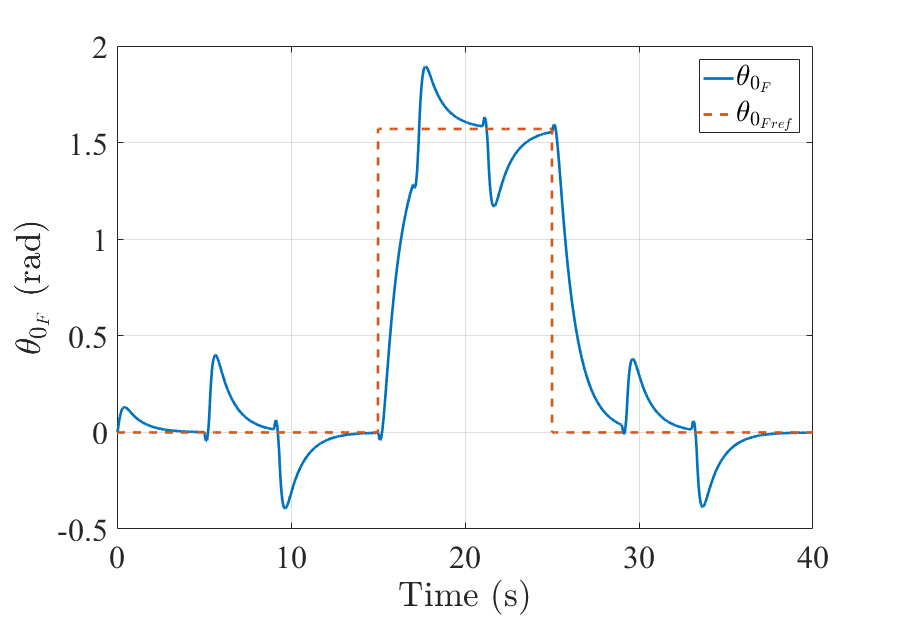}
         \caption{}
     \end{subfigure}
     \begin{subfigure}[b]{0.23\textwidth}
         \centering
         \includegraphics[trim=10 5 40 15, clip,width=\textwidth]{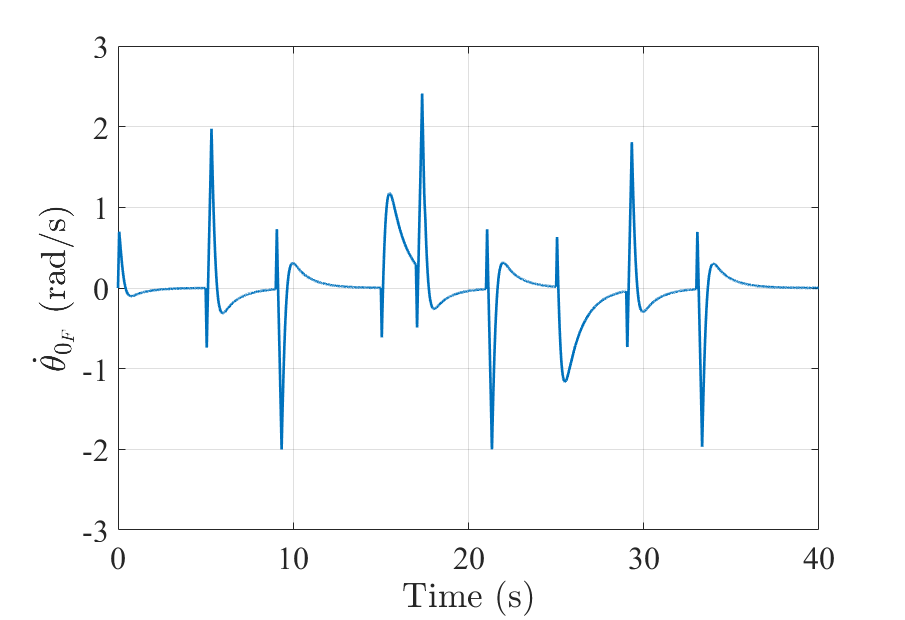}
         \caption{}
     \end{subfigure}
     \begin{subfigure}[b]{0.23\textwidth}
         \centering
         \includegraphics[trim=3 5 47 15, clip,width=\textwidth]{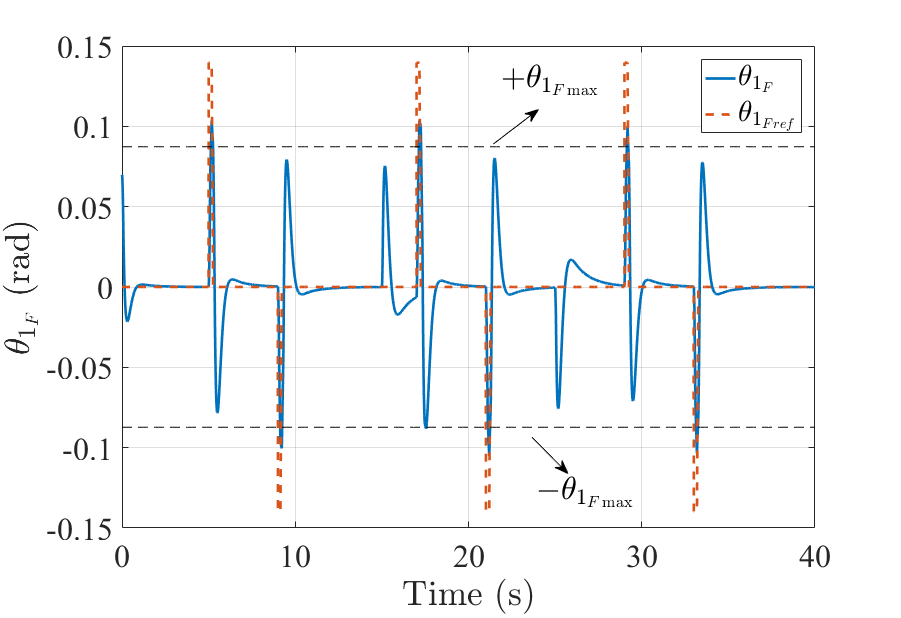}
         \caption{}
    \end{subfigure}
         \begin{subfigure}[b]{0.23\textwidth}
         \centering
         \includegraphics[trim=3 5 47 15, clip,width=\textwidth]{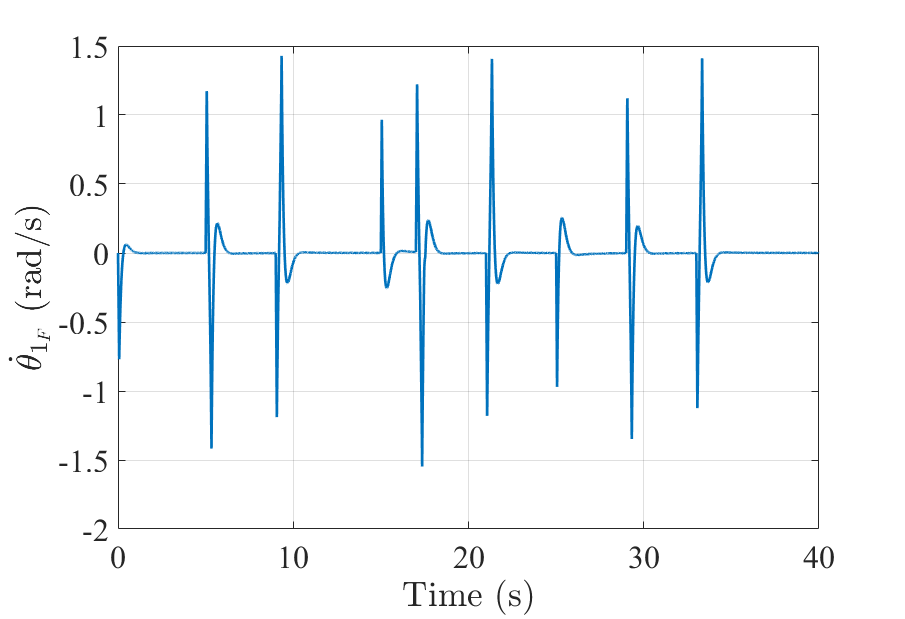}
         \caption{}
    \end{subfigure}
         \begin{subfigure}[b]{0.23\textwidth}
         \centering
         \includegraphics[trim=10 5 40 15, clip,width=\textwidth]{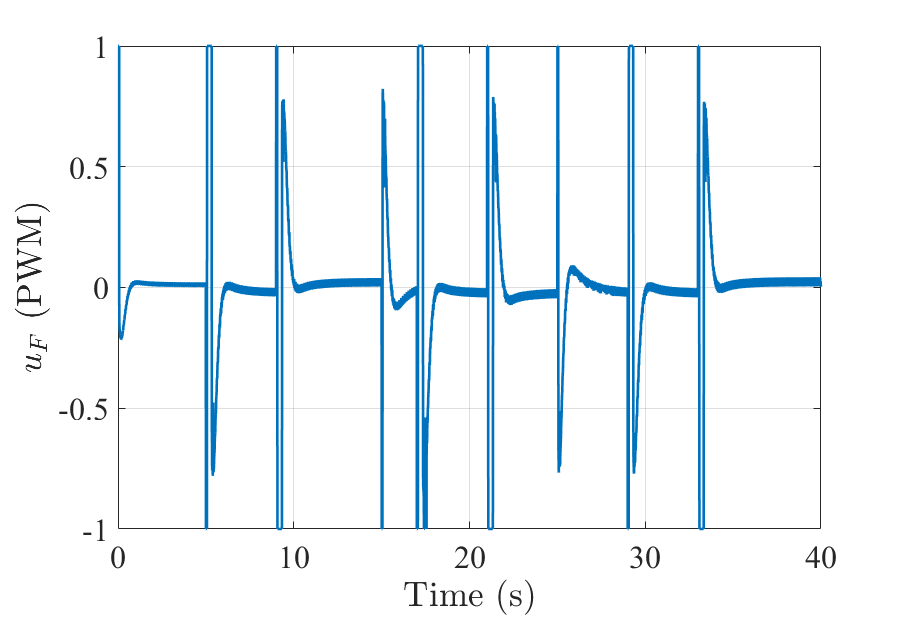}
         \caption{}
    \end{subfigure}
         \begin{subfigure}[b]{0.23\textwidth}
         \centering
         \includegraphics[trim=10 5 40 5, clip,width=\textwidth]{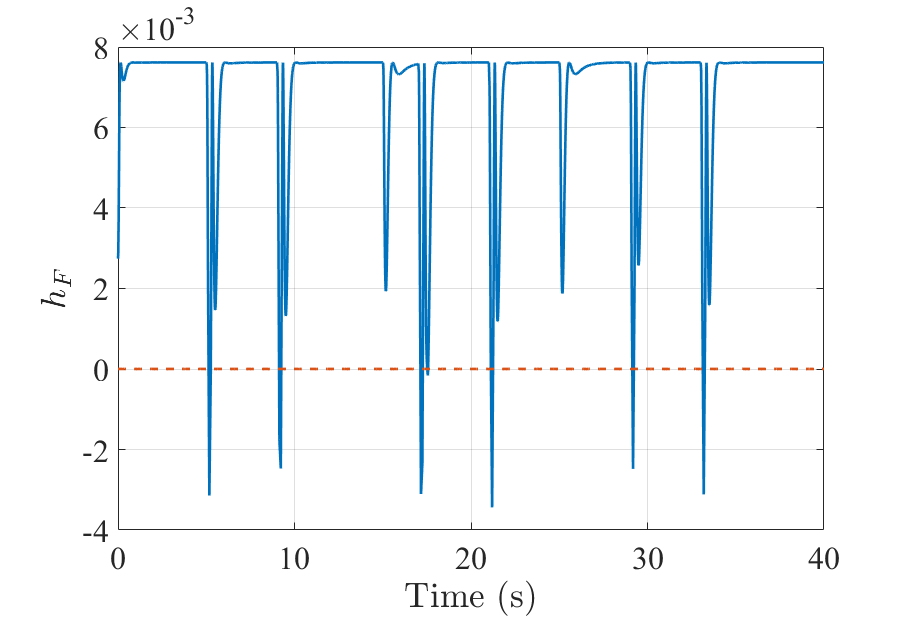}
         \caption{}
    \end{subfigure}
        \caption{Numerical simulation (Furuta pendulum) - LQR with ECBF considering real dynamics.}
        \label{fig:FUR_SIM_3}
\end{figure}

The QP is implemented using Hildreth's QP procedure \cite{Hildreth}, which is solved in polynomial time. We set $K_{b_{F}}=\left [ 3000\;\;180 \right ]$ for the pole placement controller $\mu _{b_{F}}$. Considering this value for $K_{b_{F}}$ and the nominal system dynamics, i.e., without increasing in the masses $m_{0_{F}}$ and $m_{1_{F}}$, the safety constraint is respected, as shown in Fig. \ref{fig:FUR_SIM_2}. It can be observed that $\left | \theta _{1_{F}} \right |$ never exceeds $\theta _{1_{F\max}}$, being within the safe set (\ref{eq:SAFE}). However, when it is considered the real system dynamics, i.e., with increasing in the masses $m_{0_{F}}$ and $m_{1_{F}}$, the safety constraint is no longer respected, as shown in Fig. \ref{fig:FUR_SIM_3}. Therefore, the safety constraint with ECBF is not robust.

\begin{figure}[h]
     \centering
     \begin{subfigure}[b]{0.23\textwidth}
         \centering
         \includegraphics[trim=10 5 40 15, clip,width=\textwidth]{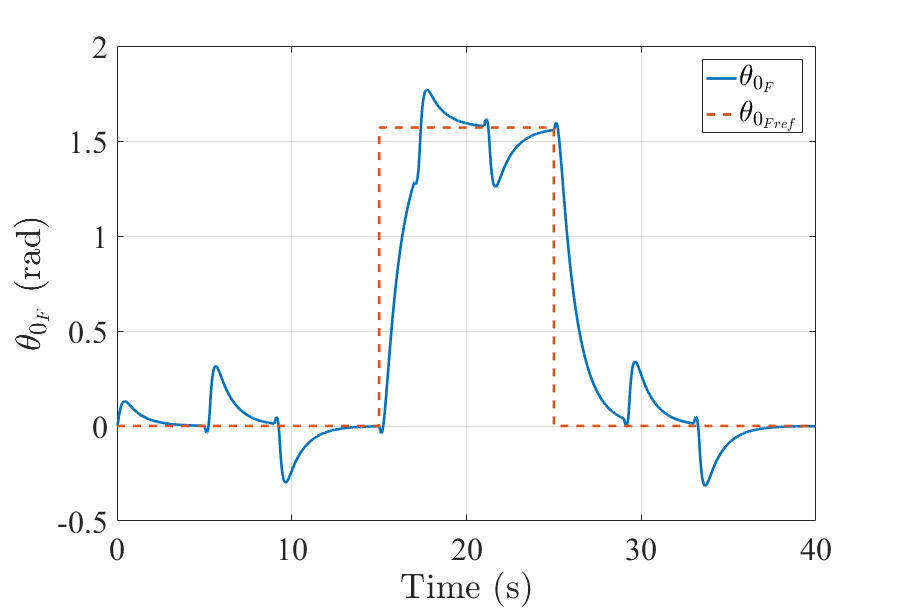}
         \caption{}
     \end{subfigure}
     \begin{subfigure}[b]{0.23\textwidth}
         \centering
         \includegraphics[trim=10 5 40 15, clip,width=\textwidth]{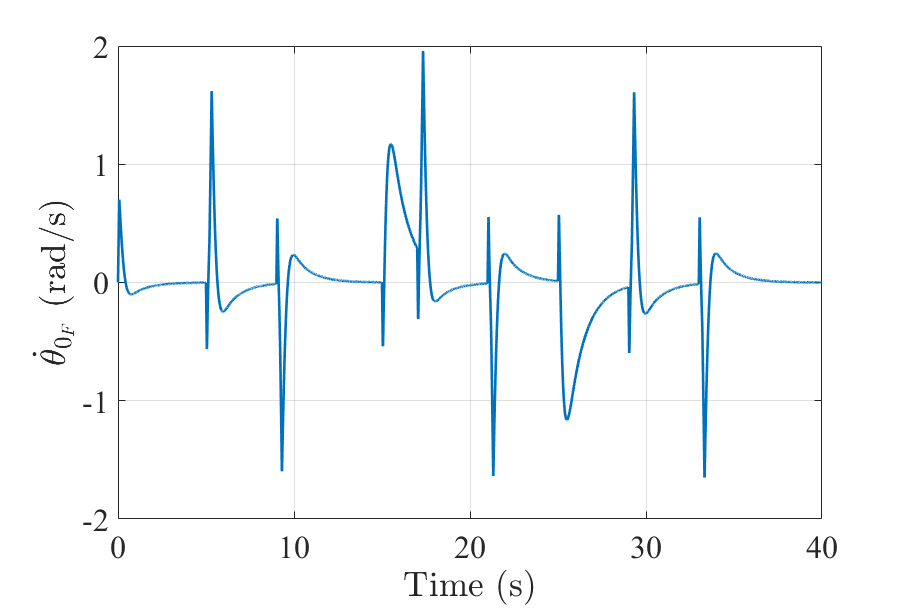}
         \caption{}
     \end{subfigure}
     \begin{subfigure}[b]{0.23\textwidth}
         \centering
         \includegraphics[trim=3 5 47 15, clip,width=\textwidth]{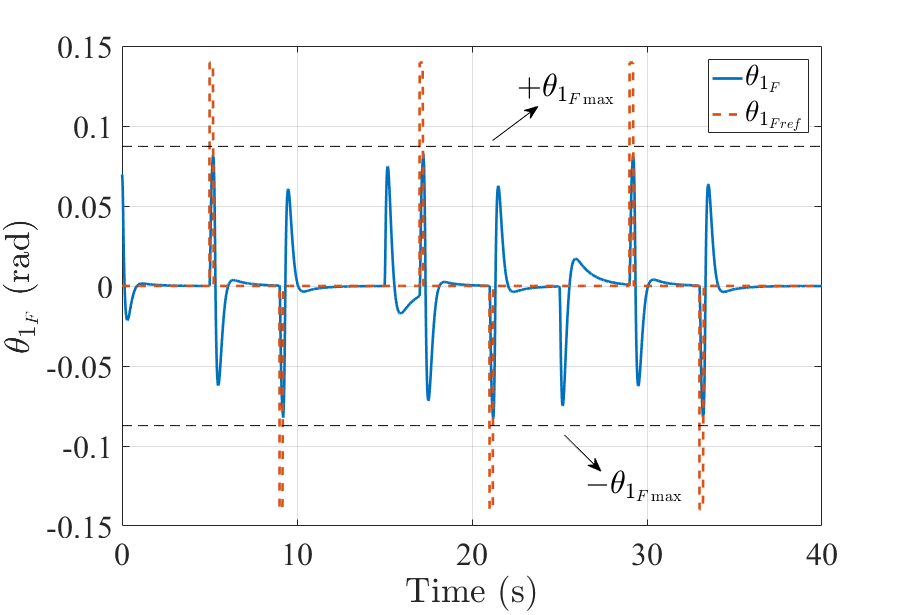}
         \caption{}
    \end{subfigure}
         \begin{subfigure}[b]{0.23\textwidth}
         \centering
         \includegraphics[trim=3 5 47 15, clip,width=\textwidth]{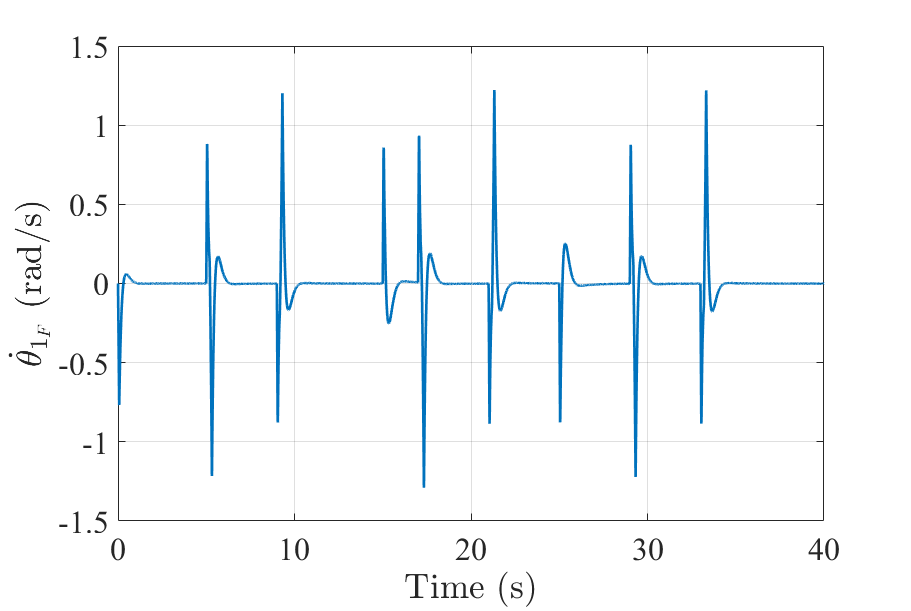}
         \caption{}
    \end{subfigure}
         \begin{subfigure}[b]{0.23\textwidth}
         \centering
         \includegraphics[trim=10 5 40 15, clip,width=\textwidth]{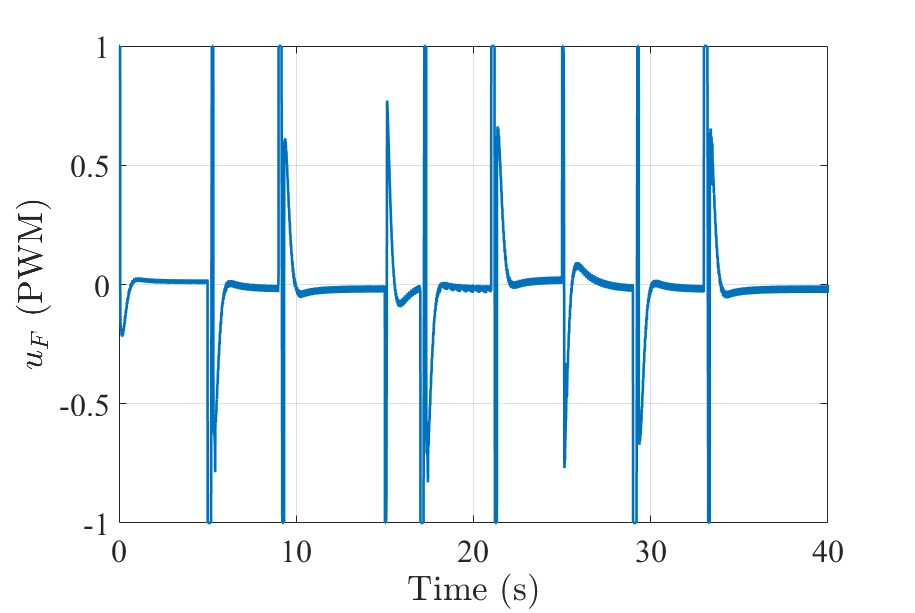}
         \caption{}
    \end{subfigure}
         \begin{subfigure}[b]{0.23\textwidth}
         \centering
         \includegraphics[trim=10 5 40 5, clip,width=\textwidth]{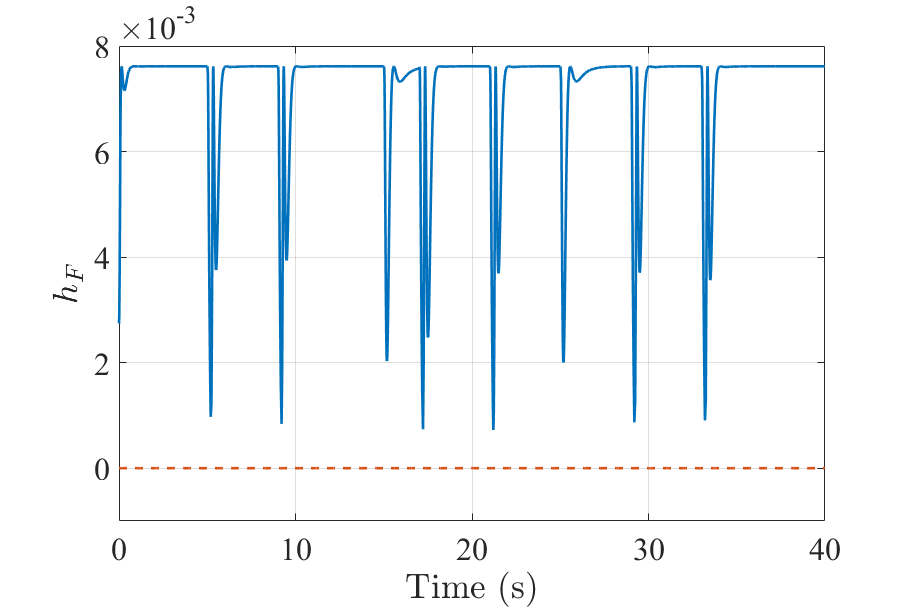}
         \caption{}
    \end{subfigure}
        \caption{Numerical simulation (Furuta pendulum) - LQR with SMCBF considering real dynamics.}
        \label{fig:FUR_SIM_4}
\end{figure}

Finally, to deal with model uncertainties, the proposed control framework (\ref{eq:RCBF13}) is applied. The design parameters are $\lambda _{F}=10$, for the sliding surface (\ref{eq:RCBF5}), $\eta _{F}=5$, for the sliding condition (\ref{eq:RCBF4}) and the boundary layer thickness $\Phi _{F}=0.1$. We empirically set $h_d = 10^{-4}$. $K_{smc_{F}}$, defined in (\ref{eq:RCBF8}), is chosen in order to satisfy (\ref{eq:RCBF10}). For the design, it was considered that $\Delta_{F_{\max}} = 1$. The simulation results are presented in Fig. \ref{fig:FUR_SIM_4}. It can be observed that $\left | \theta _{1_{F}} \right |$ never exceeds $\theta _{1_{F\max}}$ and the SMCBF $h_{F}$ respects the safe set (\ref{eq:SAFE}). Therefore, with the proposed SMCBF, the safety constraint is robust.

\section{MAGLEV}
\label{sec:MAG}

This section presents the modeling and the numerical results of the MAGLEV with the proposed control framework.

\subsection{System Modeling}

\begin{figure}[h]
\begin{center}
\includegraphics[width=6cm]{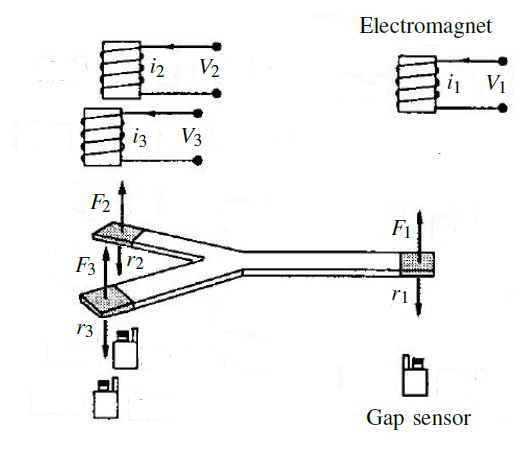}
\caption{Schematic diagram of the MAGLEV \cite{Tsujino}.}
\label{fig:maglev1}
\end{center}
\end{figure}

The MAGLEV system analyzed in this work is based on the experimental apparatus described in \cite{Tsujino}. The system is nonlinear, open loop unstable and multiple-input-multiple-output (MIMO). The schematic diagram of the MAGLEV is presented in Fig. \ref{fig:maglev1}. The system is constituted by a Y shape plate made of aluminum with small pieces of iron mounted at the edges and that must be levitated by electromagnetic forces. The inputs are represented by attractive forces $F_{1}$, $F_{2}$ and $F_{3}$ generated from three electromagnets. The controller provides voltage command signals $V_{1}$, $V_{2}$ and $V_{3}$ that are converted to proportional current signals $i_{1}$, $i_{2}$ and $i_{3}$ by power amplifiers in order to generate the corresponding attractive forces. The outputs are represented by three plate positions $r_{1}$, $r_{2}$ and $r_{3}$, measured by gap sensors mounted below the edges of the plate.

\begin{figure}[h]
\begin{center}
\includegraphics[width=6cm]{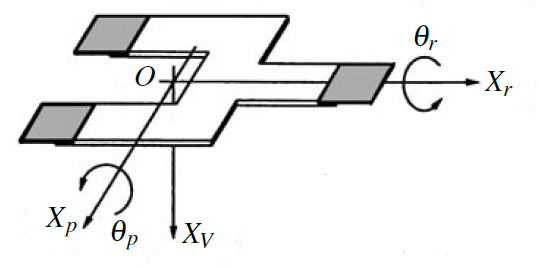}
\caption{Coordinate axis of the plate \cite{Tsujino}.}
\label{fig:maglev2}
\end{center}
\end{figure}

\begin{figure}[h]
\begin{center}
\includegraphics[width=6cm]{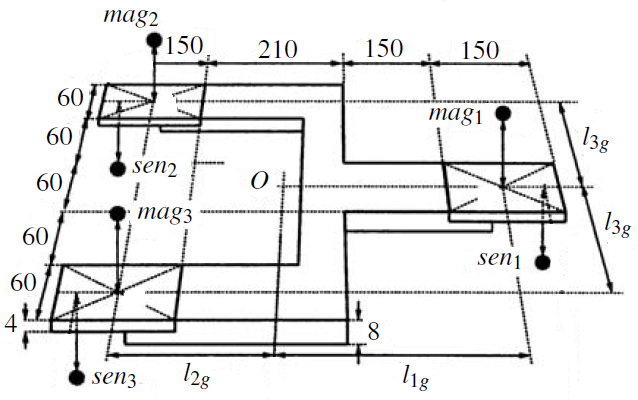}
\caption{Parameters of the MAGLEV \cite{Tsujino}.}
\label{fig:maglev3}
\end{center}
\end{figure}

The coordinate axis of the plate $X_{V}$, $X_{p}$ and $X_{r}$ are presented in Fig. \ref{fig:maglev2}. $x_{v}$ is the vertical gap length between the electromagnet and the plate at the origin $O$, right above the center of gravity $G$, while $\theta_{p}$ and $\theta_{r}$ are the pitching and rotating angles, respectively. The parameters of the MAGLEV are the mass of the plate $M$, the moments of inertia around the origin $O$ in pitching direction $J_{pm}$ and in rolling direction $J_{rm}$, and the constants $k_{1}$, $k_{2}$ and $k_{3}$ related to each electromagnet \cite{Tsujino}. Other parameters can be seen directly in the Fig. \ref{fig:maglev3}.

The equations of vertical, pitching, and rotating motions can be described respectively as \cite{Tsujino}
\begin{equation} \label{eq:maglev1}
M\ddot{x}_{v}=Mg-(F_{1}+F_{2}+F_{3}),
\end{equation}
\begin{equation} \label{eq:maglev2}
J_{pm}\ddot{\theta }_{p}=F_{1}l_{1g}-(F_{2}+F_{3})l_{2g}-Mgd_{ml}\sin \theta _{p},
\end{equation}
\begin{equation} \label{eq:maglev3}
J_{rm}\ddot{\theta }_{r}=(F_{2}-F_{3})l_{3g}-Mgd_{ml}\sin \theta _{r},
\end{equation}
where $d_{ml}$ is the distance between the origin $O$ and the center of gravity $G$, and $g$ is the gravitational acceleration.

The plate positions $r_{1}$, $r_{2}$ and $r_{3}$ have the same directions as $x_{v}$ and are given by \cite{Tsujino}
\begin{equation} \label{eq:maglev4}
r_{1}=x_{v}-l_{1g}\tan \theta _{p},
\end{equation}
\begin{equation} \label{eq:maglev5}
r_{2}=x_{v}+l_{2g}\tan \theta _{p}-l_{3g}\tan \theta _{r},
\end{equation}
\begin{equation} \label{eq:maglev6}
r_{3}=x_{v}+l_{2g}\tan \theta _{p}+l_{3g}\tan \theta _{r},
\end{equation}
and the electromagnets attractive forces can be written as a nonlinear function of the input voltages $V_{j}$ and plate positions $r_{j}$, such that \cite{Tsujino}:
\begin{equation} \label{eq:maglev7}
F_{j}:=k_{j}\left ( \frac{V_{j}}{r_{j}} \right )^{2}\;\;j=1,2,3.
\end{equation}

The system can be represented by:
\begin{equation} \label{eq:maglev8}
\dot{x}_{ml}=f_{ml}(x_{ml})+g_{ml}(x_{ml})u_{ml},
\end{equation}
\begin{equation} \label{eq:maglev9}
y_{ml}=o_{ml}(x_{ml}),
\end{equation}
where $x_{ml} = \left [x_{v}\;\;\theta _{p}\;\;\theta _{r} \;\;\dot{x}_{v}\;\;\dot{\theta}_{p}\;\;\dot{\theta}_{r}\right]^{T}$ is the state vector, $u_{ml}=\left [ F_{1}\;\;F_{2}\;\;F_{3} \right ]^{T}$ is the input vector, $y_{ml}=\left [ r_{1}\;\;r_{2}\;\;r_{3} \right ]^{T}$ is the output vector, and
\begin{equation} \label{eq:maglev10}
f_{ml}(x_{ml})= \left [ \dot{x}_{v}\;\;\dot{\theta}_{p}\;\;\dot{\theta}_{r}\;\;g\;\;\frac{-Mgd\sin \theta _{p}}{J_{pm}}\;\;\frac{-Mgd\sin \theta _{r}}{J_{rm}} \right ]^{T},
\end{equation}
\begin{equation} \label{eq:11}
g_{ml}(x_{ml}) = \left [ \begin{array}{cccccc}0&0&0\\0&0&0\\0&0&0\\-\frac{1}{M} & -\frac{1}{M} & -\frac{1}{M}\\ \frac{l_{1g}}{J_{pm}} & -\frac{l_{2g}}{J_{pm}} & -\frac{l_{2g}}{J_{pm}}\\0 & \frac{l_{3g}}{J_{rm}} & -\frac{l_{3g}}{J_{rm}}\\ \end{array} \right ],
\end{equation}
\begin{equation} \label{eq:maglev12}
o_{ml}(x_{ml}) = \left [ \begin{array}{ll}x_{v}-l_{1g}\tan \theta _{p}\\x_{v}+l_{2g}\tan \theta _{p}-l_{3g}\tan \theta _{r}\\x_{v}+l_{2g}\tan \theta _{p}+l_{3g}\tan \theta _{r} \\\end{array} \right ].
\end{equation}

\subsection{Nominal Control - SMC}

A nominal control law $u_{no_{ml}}$ must be designed so that the plate positions $y_{ml}=\left [ r_{1}\;\;r_{2}\;\;r_{3} \right ]^{T}$ track the reference inputs $y_{ml_{d}}=\left [ r_{1_{d}}\;\;r_{2_{d}}\;\;r_{3_{d}} \right ]^{T}$. As the system is nonlinear and present model uncertainties, we apply SMC for tracking \cite{Slotine}, \cite{Khalil}, such as in section \ref{sec:RCBF} for safety.

To generate a direct relationship between the output $y_{ml}$ and the input $u_{no_{ml}}$, the output must be differentiated twice, such that:

\begin{equation} \label{eq:SMC1}
\ddot{y}_{ml}=f_{y_{ml}}(x_{ml})+g_{y_{ml}}(x_{ml})u_{no_{ml}},
\end{equation}
where $f_{y_{ml}}(x_{ml})$ and $g_{y_{ml}}(x_{ml})$ are nonlinear functions of the state. It is considered that $f_{y_{ml}}(x_{ml})$ and $g_{y_{ml}}(x_{ml})$ in (\ref{eq:SMC1}) represent the real dynamics and are unknown. However, the controller's design is based on the nominal dynamics $\bar{f}_{y_{ml}}(x_{ml})$ and $\bar{g}_{y_{ml}}(x_{ml})$.

We consider a time-varying sliding surface $S_{ml_{c}}(y_{ml},t)$ given by
\begin{equation} \label{eq:SMC2}
S_{ml_{c}}=\dot{\tilde{y}}_{ml}+\lambda_{ml_{c}}\tilde{y}_{ml},
\end{equation}
where $\tilde{y}_{ml}=y_{ml}-y_{ml_{d}}$, $\lambda_{ml_{c}}$ is a strictly positive constant and using (\ref{eq:SMC1})
\begin{equation} \label{eq:SMC3}
\begin{array}{ll}
\dot{S}_{ml_{c}}=\ddot{y}_{ml}-\ddot{y}_{ml_{d}}+\lambda_{ml_{c}}\dot{\tilde{y}}_{ml}\\
=f_{y_{ml}}+g_{y_{ml}}u_{no_{ml}}-\ddot{y}_{ml_{d}}+\lambda_{ml_{c}}\dot{\tilde{y}}_{ml}. \\
\end{array}
\end{equation}

The equivalent control $\bar{u}_{no_{ml}}$ designed using the nominal dynamics and that would achieve $\dot{S}_{ml_{c}}=0$ is given by
\begin{equation} \label{eq:SMC4}
\bar{u}_{no_{ml}}=\bar{g}_{y_{ml}}^{-1}\left [ -\bar{f}_{y_{ml}}+\ddot{y}_{ml_{d}}-\lambda_{ml_{c}}\dot{\tilde{y}}_{ml} \right ].
\end{equation}

In order to satisfy the sliding condition despite uncertainties on the dynamics and considering a boundary layer to avoid chattering, we add to $\bar{u}_{no_{ml}}$ a term discontinuous across the surface $S_{ml_{c}}=0$:
\begin{equation} \label{eq:SMC5}
{u}_{no_{ml}}=\bar{u}_{no_{ml}}-\bar{g}_{y_{ml}}^{-1}K_{ml_{c}}\rm sat(S_{ml_{c}}/\Phi_{ml_{c}}),
\end{equation}
where $\Phi_{ml_{c}}$ is the boundary layer thickness.

The gain $K_{ml_{c}}$ must satisfy the sliding mode condition (\ref{eq:RCBF4}) and it is given by
\begin{equation} \label{eq:SMC6}
\begin{array}{ll}
K_{ml_{c}}\geq(g_{y_{ml}}^{-1}\bar{g}_{y_{ml}})(\eta_{ml_{c}}+f_{y_{ml}})-\bar{f}_{y_{ml}}
\\
+(I-g_{y_{ml}}^{-1}\bar{g}_{y_{ml}})(\ddot{y}_{ml_{d}}-\lambda_{ml_{c}}\dot{\tilde{y}}_{ml}),
\end{array}
\end{equation}
where $\eta_{ml_{c}}$ is a strictly positive constant shown in (\ref{eq:RCBF4}).




\subsection{Numerical Results}

The behavior of the MAGLEV with the proposed control framework is verified through numerical simulations with MATLAB/Simulink. The numerical values of the parameters, described in \cite{Tsujino}, are $l_{1g}=0.306\;\rm m$, $l_{2g}=0.203\;\rm m$, $l_{3g}=0.120\;\rm m$, $M=1.93\;\rm Kg$, $g=9.81\;\rm m/s^{2}$, $J_{pm}=6.43\times 10^{-2}\;\rm kgm^{2}$, $J_{rm}=1.82\times 10^{-2}\;\rm kgm^{2}$, $k_{1}=3.70\times 10^{-4}\;\rm Nm^{2}/V$, $k_{2}=1.03\times 10^{-4}\;\rm Nm^{2}/V$, $k_{3}=1.36\times 10^{-4}\;\rm Nm^{2}/V$ and $d_{ml}=3.24\times 10^{-3}\;\rm m$.

Several works are proposed to satisfy a tracking objective in the MAGLEV, i.e, to track the reference inputs, but safety constraints are not considered. So, it is applied the control framework described in this work to simultaneously satisfy tracking objectives and safety constraints. SMC is applied as a nominal control law $u_{no_{ml}}$, defined in (\ref{eq:SMC5}), for tracking the reference inputs. The safety constraints are considered to guarantee that the plate positions $r_{1}$, $r_{2}$ and $r_{3}$ never exceed predetermined values $r_{1_{\max}}$, $r_{2_{\max}}$ and $r_{3_{\max}}$ and satisfy the safe set (\ref{eq:SAFE}). It is considered $r_{1_{\max}}=0.01\;\rm m$ related to $r_{1}=-0.05\;\rm m$, $r_{2_{\max}}=0.01\;\rm m$ related to $r_{2}=-0.07\;\rm m$ and $r_{3_{\max}}=0.01\;\rm m$ related to $r_{3}=-0.09\;\rm m$. The mass of the plate $M$ is increased 30\%.

The design parameters for the nominal control $u_{no_{ml}}$ are
\begin{equation} \label{eq:MAGRES1}
\lambda _{ml_{c}} = \left [ \begin{array}{ccc}50 & 0 & 0\\0 & 50 & 0\\0 & 0 & 50 \end{array} \right ], \nonumber
\end{equation}
for the sliding surface (\ref{eq:SMC2}), $\eta _{ml_{c}}= \left [ 30\;\;30\;\;30\right ]^{T}$ for the sliding condition (\ref{eq:RCBF4}) and $\Phi _{ml_{c}}= \left [ 0.05\;\;0.05\;\;0.05\right ]^{T}$ for the boundary layer thickness. The gain $K_{ml_{c}}$ is set in order to satisfy (\ref{eq:SMC6}).

\begin{figure}[h]
     \centering
     \begin{subfigure}[b]{0.32\textwidth}
         \centering
         \includegraphics[width=\textwidth]{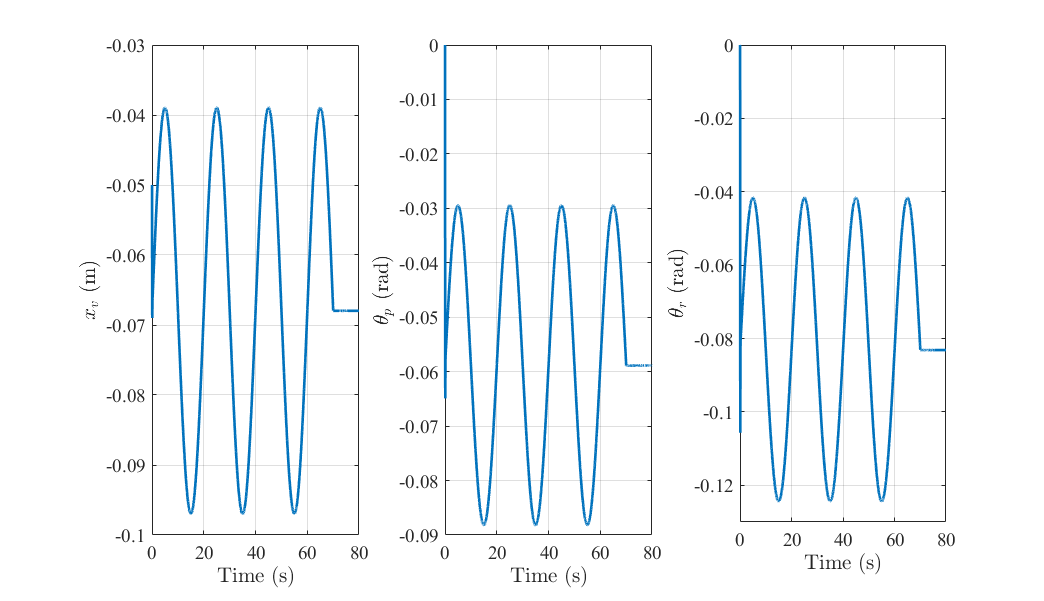}
         \caption{}
     \end{subfigure}
     \hfill
     \begin{subfigure}[b]{0.32\textwidth}
         \centering
         \includegraphics[width=\textwidth]{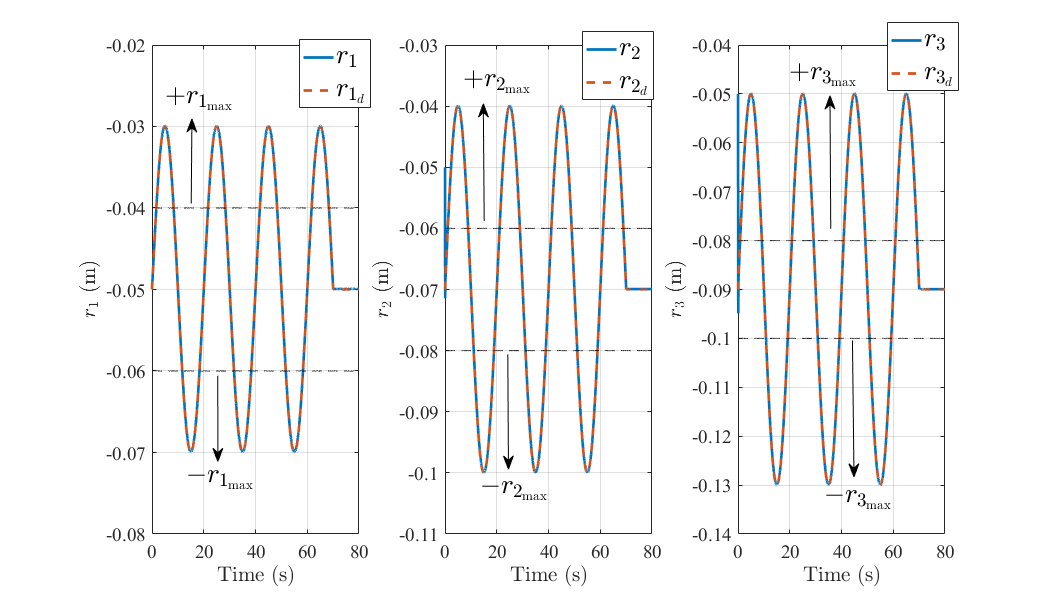}
         \caption{}
     \end{subfigure}
     \hfill
     \begin{subfigure}[b]{0.32\textwidth}
         \centering
         \includegraphics[width=\textwidth]{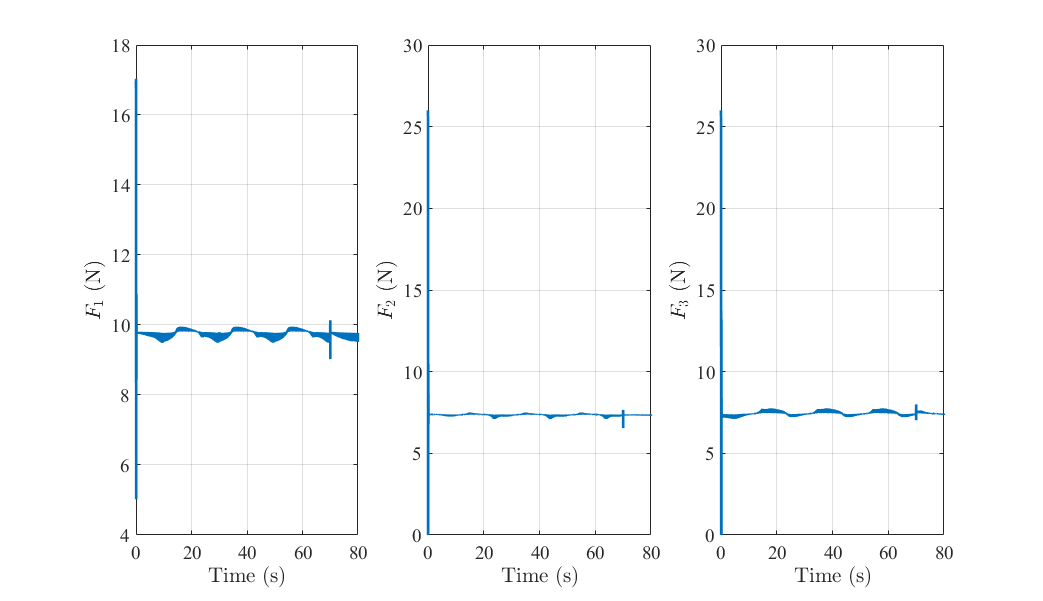}
         \caption{}
     \end{subfigure}
        \caption{Numerical simulation (MAGLEV) - SMC without ECBF.}
        \label{fig:MAGLEV_SIM_1}
\end{figure}

It is proposed an experiment whereby the plate positions $r_{1}$, $r_{2}$ and $r_{3}$ should track the reference inputs $r_{1_{d}}$, $r_{2_{d}}$ and $r_{3_{d}}$ and the safety constraints must be respected. Initially, only the SMC is applied as the nominal control law and the safety constraints are not considered. The simulation results are presented in Fig. \ref{fig:MAGLEV_SIM_1}. In all numerical simulations, the MAGLEV is assumed to start at an initial position $r_{1_{0}}=0.05\;\rm m$, $r_{2_{0}}=0.05\;\rm m$ and $r_{3_{0}}=0.05\;\rm m$. The results show that the SMC is able to track the reference inputs even with increasing in the mass of the plate $M$.

Posteriorly, the safety constraints are considered to guarantee that the plate positions $r_{1}$, $r_{2}$ and $r_{3}$ never exceed predetermined values $r_{1_{\max}}$, $r_{2_{\max}}$ and $r_{3_{\max}}$ and satisfy the safe set (\ref{eq:SAFE}); thus, the QP-based controller (\ref{eq:ECBF7}) that unifies the nominal SMC $u_{no_{ml}}$ and the safety constraints is applied. The controller is adapted to multiple safety constraints, such as
\begin{multline} \label{eq:ECBF7a}
u_{ml}^{*}(x_{ml})=\underset{u=(u_{ml},\mu_{b_{ml}})\in {\mathbb{R}}^{m+3}}{\arg\min}\;u_{ml}^{T}u_{ml}-2u_{no_{ml}}^{T}u_{ml} \\
s.t.\;\left [L_{g}L_{f}^{r-1}h_{1}(x_{ml})\;L_{g}L_{f}^{r-1}h_{2}(x_{ml})\;L_{g}L_{f}^{r-1}h_{3}(x_{ml})  \right ]u \hfill \hfill \\
+\left [ L_{f}^{r}h_{1}(x_{ml})\;L_{f}^{r}h_{2}(x_{ml})\;L_{f}^{r}h_{3}(x_{ml}) \right ]^{T}=\mu_{b_{ml}}, \\
\;\;\;\;\;\;\; \mu_{b_{ml}}=\left [ \mu_{b_{1}}\;\mu_{b_{2}}\;\mu_{b_{3}} \right ]^{T}\geq \hfill \\
-\left [ K_{b_{1}}\;K_{b_{2}}\;K_{b_{3}} \right ]\left [ \eta _{b_{1}}(x_{ml})\;\eta _{b_{2}}(x_{ml})\;\eta _{b_{3}}(x_{ml}) \right ]^{T},
\end{multline}
where we consider the following relative-degree two ($r=2$) safety constraints, expressed as the CBFs
\begin{equation} \label{eq:MAGRES4}
h_{j}(x_{ml})=(r_{j_{\max}})^{2}-(r_{j}-r_{jd})^{2},\;\;j=1,2,3.
\end{equation}

\begin{figure}
     \centering
     \begin{subfigure}[b]{0.32\textwidth}
         \centering
         \includegraphics[width=\textwidth]{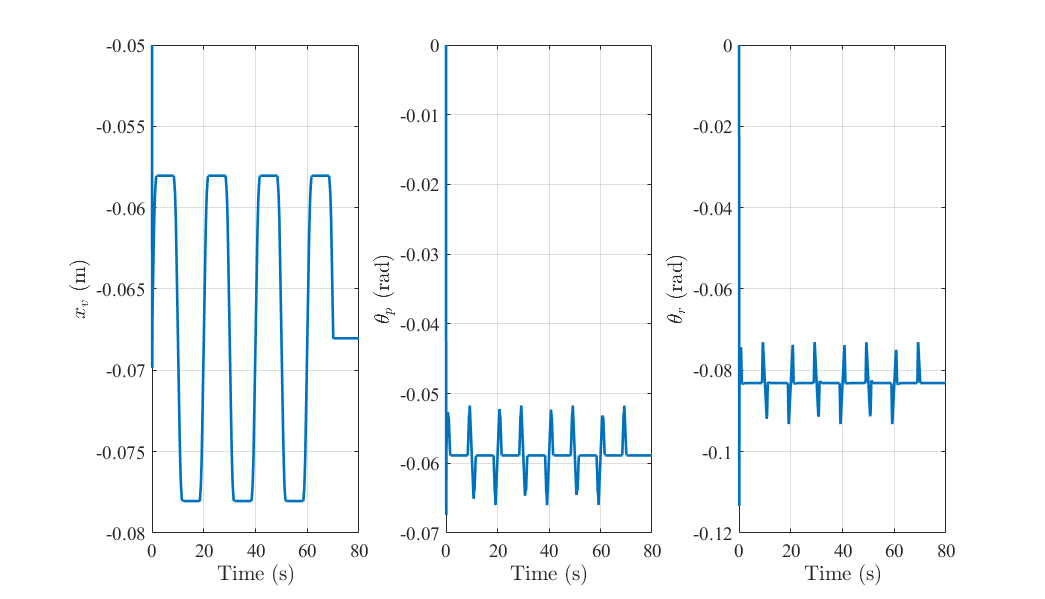}
         \caption{}
     \end{subfigure}
     \hfill
     \begin{subfigure}[b]{0.32\textwidth}
         \centering
         \includegraphics[width=\textwidth]{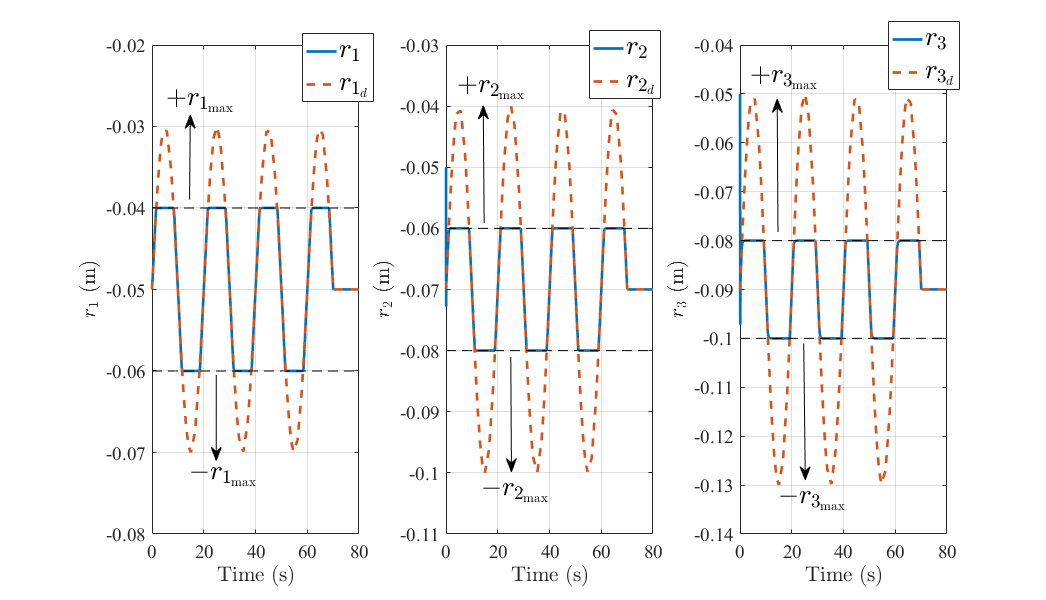}
         \caption{}
     \end{subfigure}
     \hfill
     \begin{subfigure}[b]{0.32\textwidth}
         \centering
         \includegraphics[width=\textwidth]{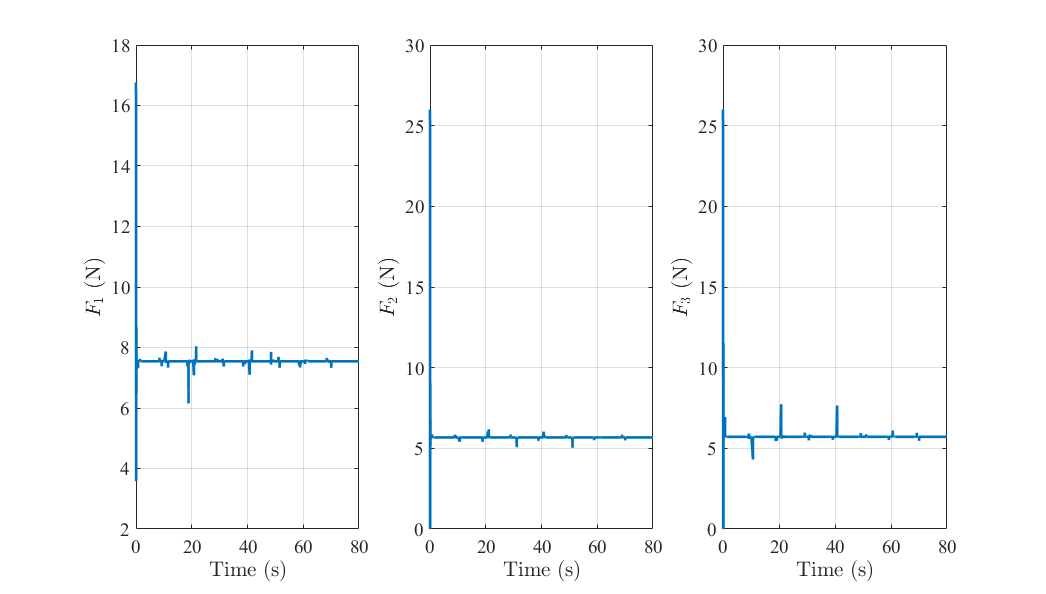}
         \caption{}
     \end{subfigure}
     \hfill
     \begin{subfigure}[b]{0.32\textwidth}
         \centering
         \includegraphics[width=\textwidth]{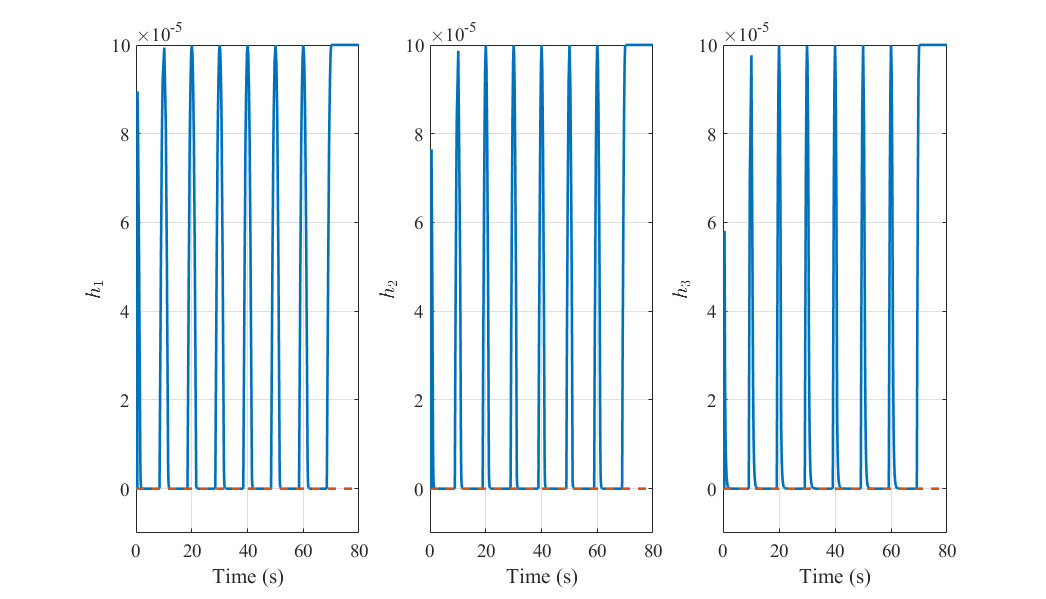}
         \caption{}
     \end{subfigure}
        \caption{Numerical simulation (MAGLEV) - SMC with ECBF considering nominal dynamics.}
        \label{fig:MAGLEV_SIM_2}
\end{figure}

\begin{figure}
     \centering
     \begin{subfigure}[b]{0.32\textwidth}
         \centering
         \includegraphics[width=\textwidth]{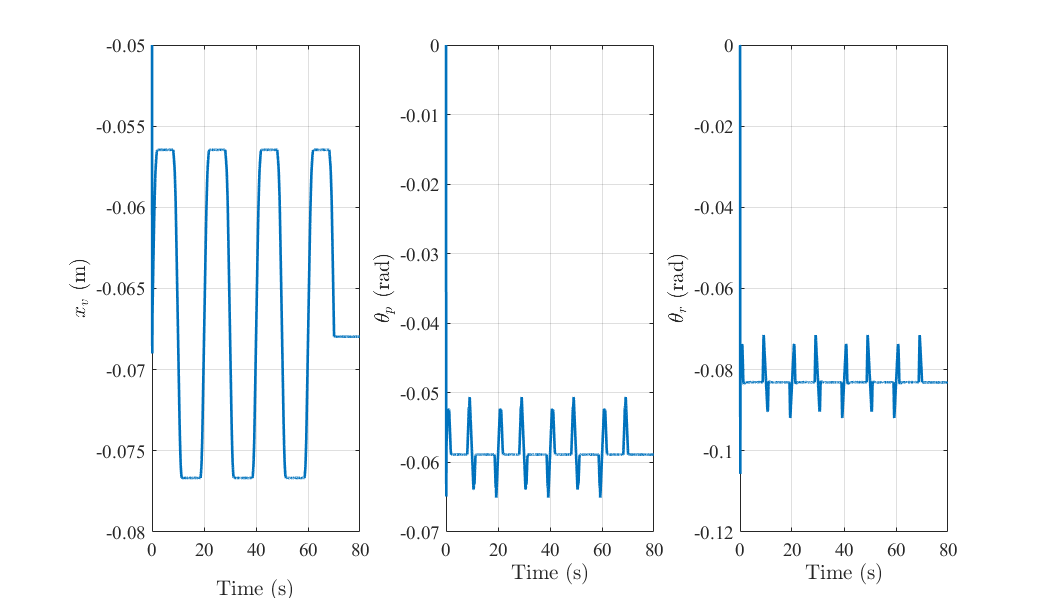}
         \caption{}
     \end{subfigure}
     \hfill
     \begin{subfigure}[b]{0.32\textwidth}
         \centering
         \includegraphics[width=\textwidth]{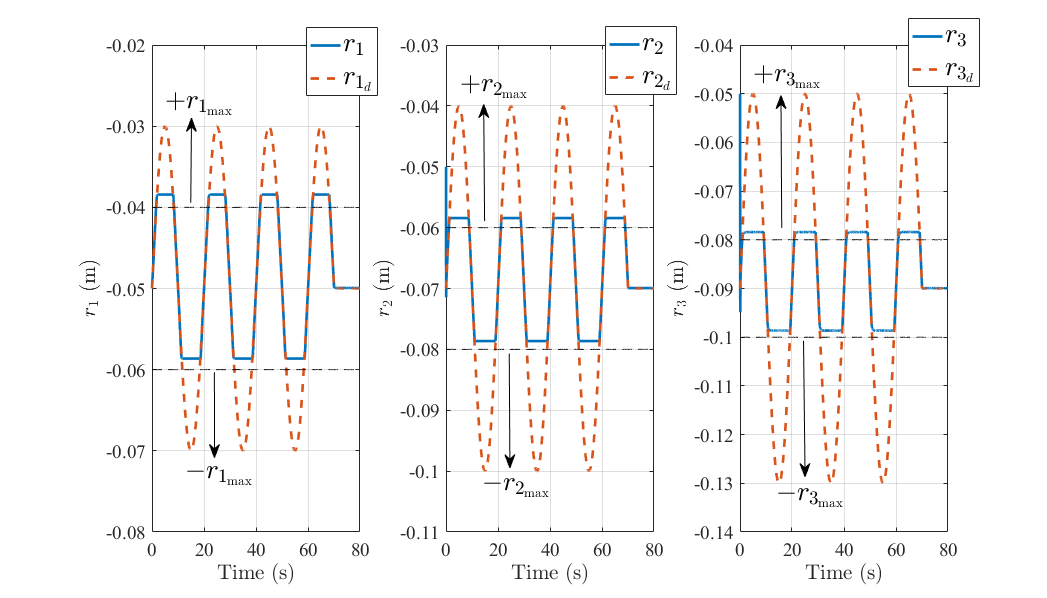}
         \caption{}
     \end{subfigure}
     \hfill
     \begin{subfigure}[b]{0.32\textwidth}
         \centering
         \includegraphics[width=\textwidth]{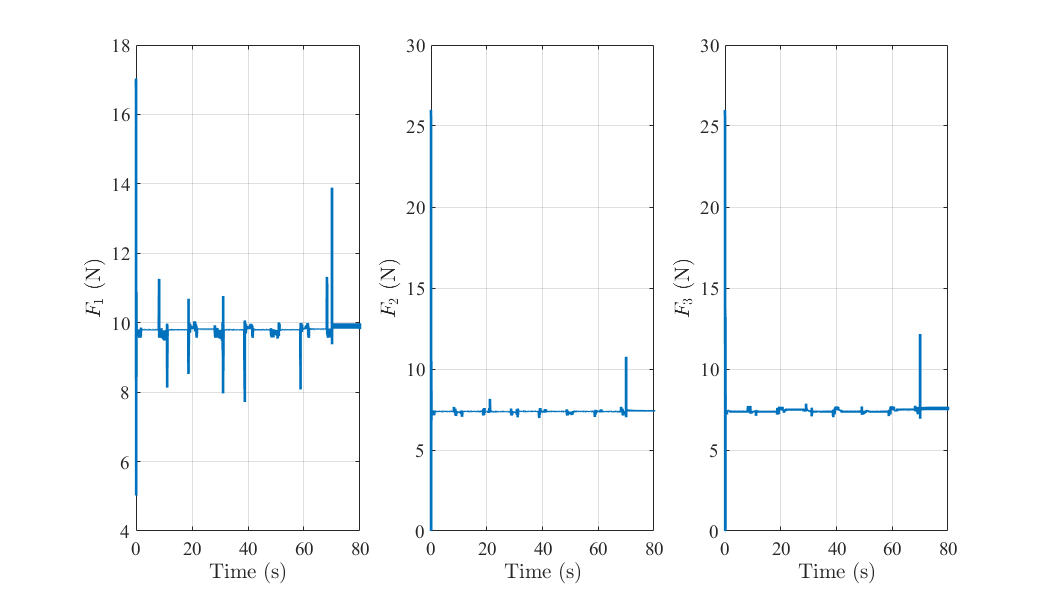}
         \caption{}
     \end{subfigure}
      \hfill
      \begin{subfigure}[b]{0.32\textwidth}
         \centering
         \includegraphics[width=\textwidth]{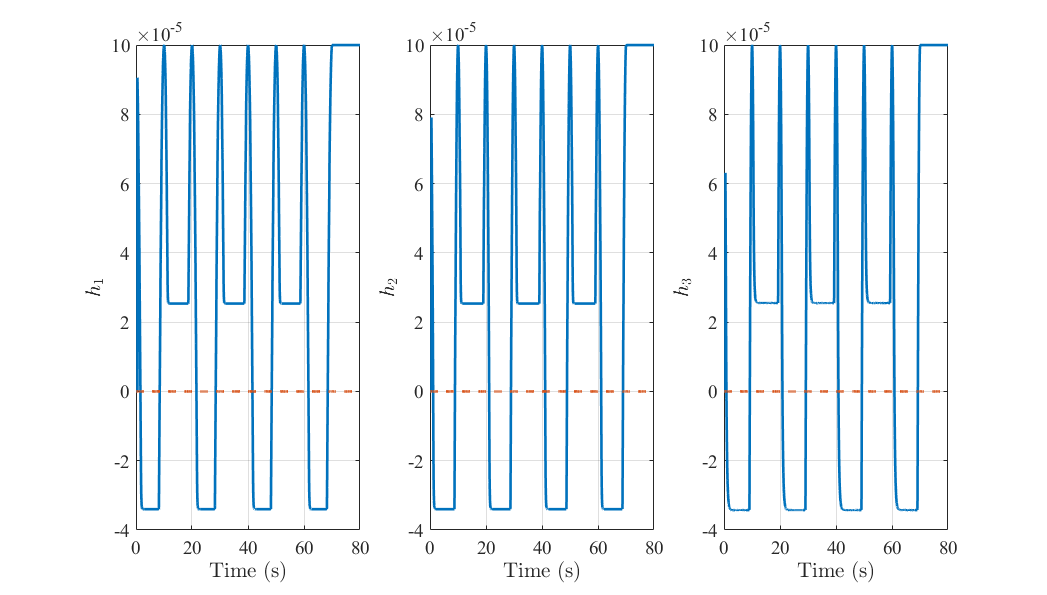}
         \caption{}
     \end{subfigure}
        \caption{Numerical simulation (MAGLEV) - SMC with ECBF considering real dynamics.}
        \label{fig:MAGLEV_SIM_3}
\end{figure}

The QP is implemented using Hildreth's QP procedure. We set $K_{b_{1}}=\left [ 2000\;\;200 \right ]$, $K_{b_{2}}=\left [ 2000\;\;200 \right ]$ and $K_{b_{3}}=\left [ 2000\;\;500 \right ]$ for the pole placement controller $\mu _{b_{ml}}$. Considering these values for $K_{b_{1}}$, $K_{b_{2}}$ and $K_{b_{3}}$, and the nominal system dynamics, i.e., without increasing in the mass of the plate $M$, the safety constraints are respected, as shown in Fig. \ref{fig:MAGLEV_SIM_2}. It can be observed that $\left | r_{j} \right |$ never exceed $r_{j_{\max}}$ and the ECBFs obtained from $h_{j}$ respect the safe set (\ref{eq:SAFE}). Initially $\left | r_{2} \right |$ and $\left | r_{3} \right |$ exceed $r_{2_{\max}}$ and $r_{3_{\max}}$, since the ECBFs were programmed to act just after the transitory due to the initial condition. However, when the mass of the plate $M$ is increased, the safety constraints are not respected, as shown in Fig. \ref{fig:MAGLEV_SIM_3}. Therefore, the ECBFs designed by pole placement are not robust.

\begin{figure}[h]
     \centering
     \begin{subfigure}[b]{0.32\textwidth}
         \centering
         \includegraphics[width=\textwidth]{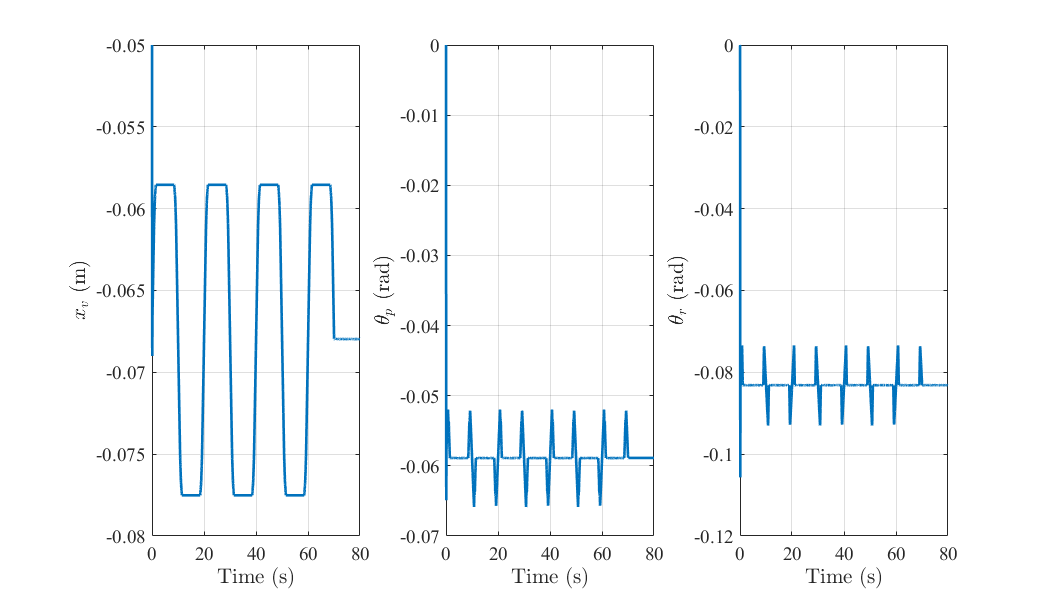}
         \caption{}
     \end{subfigure}
     \hfill
     \begin{subfigure}[b]{0.32\textwidth}
         \centering
         \includegraphics[width=\textwidth]{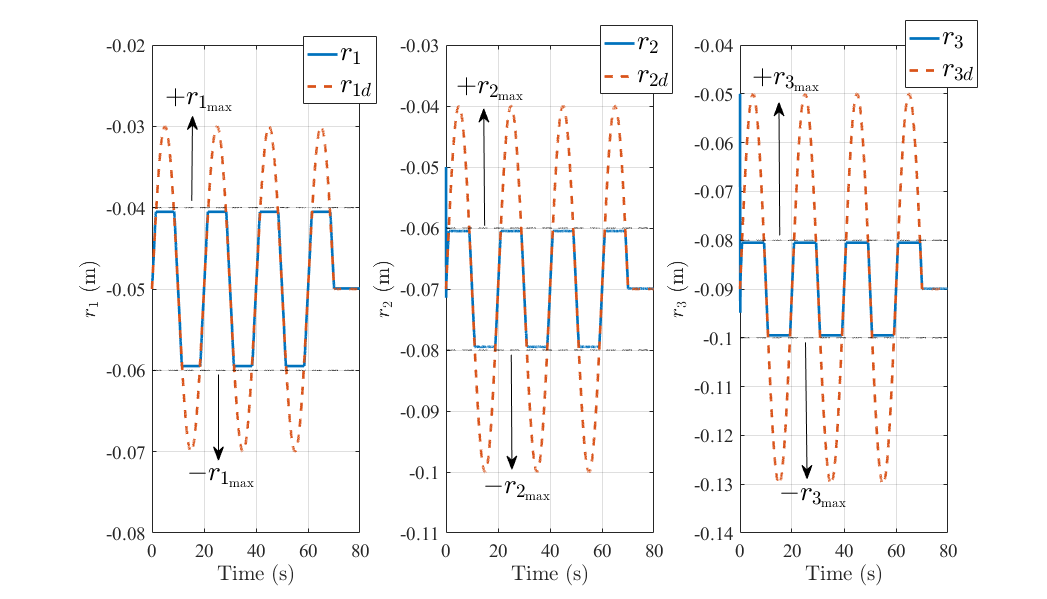}
         \caption{}
     \end{subfigure}
     \hfill
     \begin{subfigure}[b]{0.32\textwidth}
         \centering
         \includegraphics[width=\textwidth]{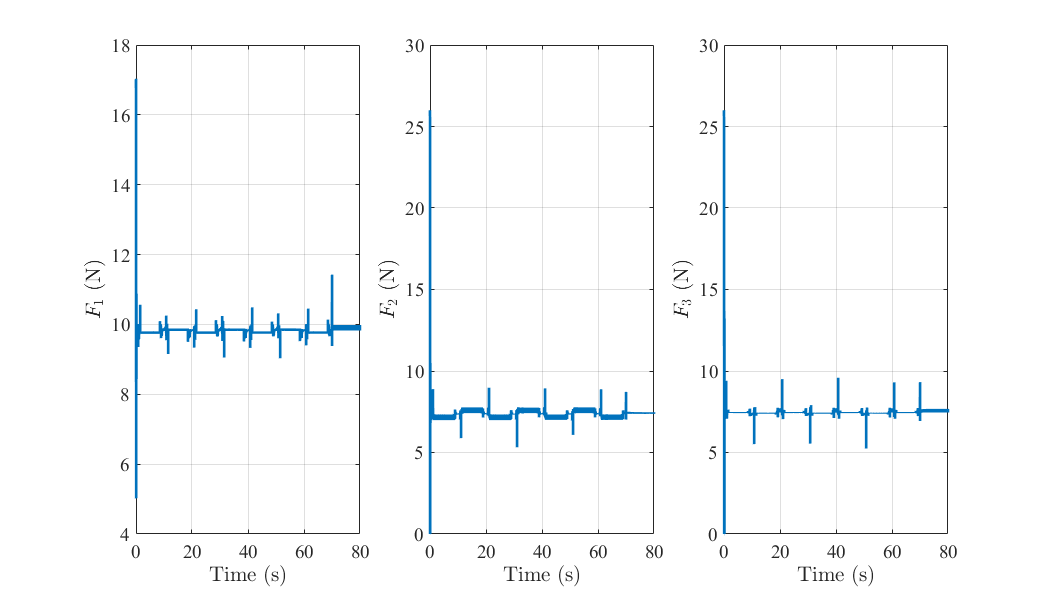}
         \caption{}
     \end{subfigure}
      \hfill
     \begin{subfigure}[b]{0.32\textwidth}
         \centering
         \includegraphics[width=\textwidth]{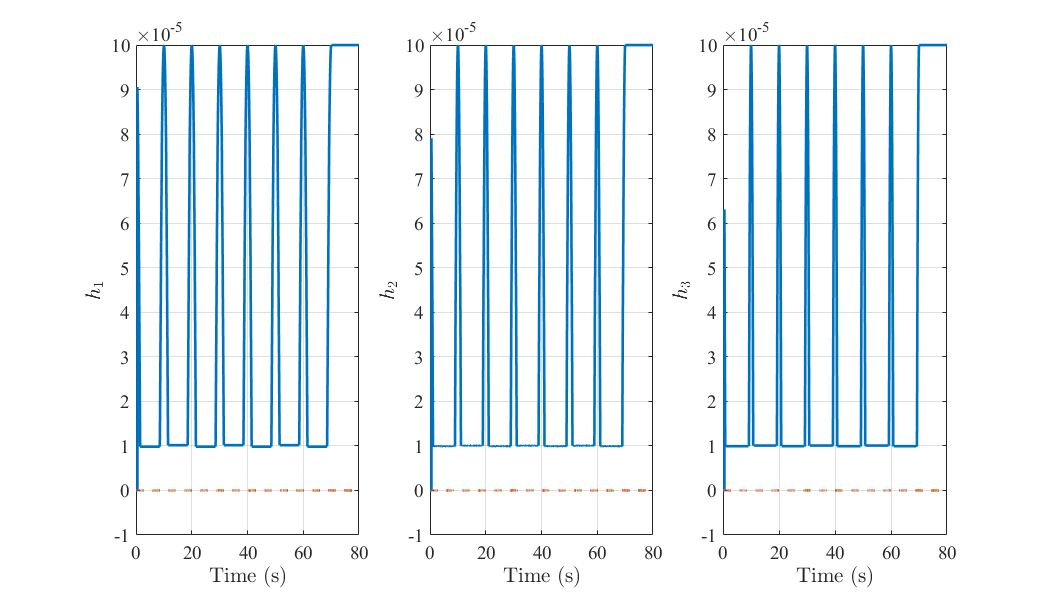}
         \caption{}
     \end{subfigure}
        \caption{Numerical simulation (MAGLEV) - SMC with SMCBF considering real dynamics.}
        \label{fig:MAGLEV_SIM_4}
\end{figure}

Finally, to deal with model uncertainties, the proposed control framework with SMCBF is considered. The controller is adapted to multiple safety constraints, such as
\begin{multline} \label{eq:RCBF13a}
u_{ml}^{*}(x_{ml})=\underset{u=(u_{ml},\mu_{b_{ml}})\in {\mathbb{R}}^{m+3}}{\arg\min}\;u_{ml}^{T}u_{ml}-2u_{no_{ml}}^{T}u_{ml} \\
s.t.\;\;\left [L_{g}L_{f}^{r-1}h_{1}(x_{ml})\;L_{g}L_{f}^{r-1}h_{2}(x_{ml})\;L_{g}L_{f}^{r-1}h_{3}(x_{ml})  \right ]u \hfill \\
+\left [ L_{f}^{r}h_{1}(x_{ml})\;L_{f}^{r}h_{2}(x_{ml})\;L_{f}^{r}h_{3}(x_{ml}) \right ]^{T}=\mu_{b_{ml}}, \\
\;\;\;\;\;\;\;\;\mu_{b_{ml}}=\left [ \mu_{b_{1}}\;\mu_{b_{2}}\;\mu_{b_{3}} \right ]^{T}\geq \left [ \bar{\mu}_{b_{1}}\;\bar{\mu}_{b_{2}}\;\bar{\mu}_{b_{3}} \right ]^{T} -  \hfill \\
K_{ml_{s}}\rm sat(S_{ml_{s}}/\Phi_{ml_{s}}),
\end{multline}
where the sliding surface $S_{ml_{s}}$ is given by (\ref{eq:RCBF5}) considering $h_{ml}=\left [ h_{1}\;\;h_{2}\;\;h_{3} \right ]^{T}$ and
\begin{equation} \label{eq:MAGRES5}
\lambda _{ml_{s}} = \left [ \begin{array}{ccc}500 & 0 & 0\\0 & 500 & 0\\0 & 0 & 500 \end{array} \right ], \nonumber
\end{equation}
for the sliding surface $S_{ml_{s}}$, $\eta _{ml_{s}}= \left [ 500\;\;500\;\;500\right ]^{T}$ for the sliding condition, $\Phi _{ml_{s}}= \left [ 0.8\;\;0.3\;\;0.3\right ]^{T}$ for the boundary layer thickness, and  $h_{d} = \left [10^{-5}\;\;10^{-5}\;\;10^{-5}\right ]^T$ (empirically set). The gain $K_{ml_{s}}$ is chosen in order to satisfy (\ref{eq:RCBF10}) and it was also assumed that $\Delta_{ml_{\max}} = \left [10\;\;10\;\;10\right ]^T$. Simulation results are presented in Fig. \ref{fig:MAGLEV_SIM_4}. It can be observed that $\left | r_{j} \right |$ never exceeds $r_{j_{\max}}$ and the SMCBFs respect the safe set (\ref{eq:SAFE}), being, therefore, robust.

\section{CONCLUSIONS}
\label{sec:CONC}

This work considers a safety-critical control framework where stability/tracking objectives, expressed as a nominal control law, and safety constraints, expressed as CBFs, are unified through QP. We proposed a SMCBF to deal with high relative-degree safety constraints and model uncertainties. The proposed scheme is numerically validated considering a Furuta pendulum and a MAGLEV. In the first case, a LQR is considered as the nominal control and a safety constraint guarantees that the pendulum angular position never exceeds a predetermined value. For the second one, a SMC is considered as a nominal control law and multiple safety constraints guarantee that the MAGLEV positions never exceed predetermined values. For both systems, we consider high relative-degree safety constraints robust against model uncertainties. The numerical results indicate that the stability/tracking objectives are reached and when the ECBFs designed by pole placement are considered, the safety constraints are respected for the nominal model, but not for the model with uncertainties, i.e., robustness is not verified. However, with the SMCBFs, the safety constraints are respected when model uncertainties are considered, i.e., robustness is verified. As suggestions of future work, we consider the validation of the proposed control framework in a physical system.

\addtolength{\textheight}{-12cm}   



\end{document}